\documentclass[10pt,a4paper,onecolumn,twoside]{article}
\usepackage[top=2cm,bottom=2cm,left=1.25cm,right=1.25cm]{geometry}

\usepackage[utf8]{inputenc}
\usepackage[T1]{fontenc}
\usepackage{adjustbox}

\usepackage{fourier}
\usepackage{amsmath}
\allowdisplaybreaks
\usepackage{booktabs}
\usepackage{graphicx}
\usepackage{marginnote}

\usepackage{physics}
\usepackage{amsfonts}
\usepackage{amssymb}
\usepackage{stmaryrd}
\usepackage{mathrsfs}
\usepackage{bm}

\usepackage{amsthm}
\theoremstyle{remark}

\theoremstyle{theorem}

\usepackage{framed}
\usepackage[hidelinks]{hyperref}

\mathcode`@="8000 
{\catcode`\@=\active\gdef@{\mkern1mu}}

\newcommand*\oline[1]{%
  \vbox{%
    \hrule height 0.5pt%
    \kern0.25ex%
    \hbox{%
      \kern-0.1em%
      \ifmmode#1\else\ensuremath{#1}\fi%
      \kern-0.1em%
    }
  }
}

\usepackage[normalem]{ulem}
\usepackage{xcolor}

\newcommand{\pdiff}[2]{\frac{\partial #1}{\partial #2}}

\title{Modelling the non-linear viscoelastic behaviour of brain tissue in torsion}
\author{Griffen Small \textsuperscript{a}, Francesca Ballatore \textsuperscript{b}, Chiara Giverso \textsuperscript{b} and Valentina Balbi \textsuperscript{a} \\ ~ \\
\emph{\textsuperscript{a}School of Mathematical and Statistical Sciences, University of Galway,}\\ \emph{College Road, Galway, H91 TK33, Ireland}
 \\ ~ \\
\emph{\textsuperscript{b}Department of Mathematical Sciences ``G. L. Lagrange'', Politecnico di Torino,}\\ \emph{Corso Duca degli Abruzzi 24, 10129
Turin, Italy}
}

\date{}

\begin{document}

\pagestyle{headings}
\thispagestyle{empty}
\maketitle

\begin{abstract}
	\noindent
Brain tissue accommodates non-linear deformations and exhibits time-dependent mechanical behaviour. The latter is one of the most pronounced features of brain tissue, manifesting itself primarily through viscoelastic effects such as stress relaxation. To investigate its viscoelastic behaviour, we performed ramp-and-hold relaxation tests in torsion on freshly slaughtered cylindrical ovine brain samples ($25\,\,\text{mm}$ diameter and $\sim 10\,\,\text{mm}$ height). The tests were conducted using a commercial rheometer at varying twist rates of $\{40,240,400\}\,\,\text{rad}\,\,\text{m}^{-1}\,\,\text{s}^{-1}$, with the twist remaining fixed at $\sim 88\,\,\text{rad}\,\,\text{m}^{-1}$, which generated two independent datasets for torque and normal force. The complete set of viscoelastic material parameters was estimated via a simultaneous fit to the analytical expressions for the torque and normal force predicted by the modified quasi-linear viscoelastic model. The model's predictions were further validated through finite element simulations in FEniCS. Our results show that the modified quasi-linear viscoelastic model---recently reappraised and largely unexploited---accurately fits the experimental data. Moreover, the estimated material parameters are in line with those obtained in previous studies on brain samples under torsion. When coupled with bespoke finite element models, these material parameters could enhance our understanding of the forces and deformations involved in traumatic brain injury and contribute to the design of improved headgear for sports such as boxing and motorsports. On the other hand, our novel testing protocol offers new insights into the mechanical behaviour of soft tissues other than the brain. \\

	
\end{abstract}


\section{Introduction}
\label{sec: Introduction}

Among all the tissues in the human body, brain tissue is the softest, with a shear modulus of the order of one kilopascal \cite{budday20}. It is also, arguably, the most important, intricate and least understood. As is the case for most biological soft tissues, brain tissue displays highly complex mechanical behaviour: it can accommodate finite deformations and its response to applied forces is markedly non-linear \cite{balbi19,destrade15}; it is incompressible and biphasic, consisting of a porous solid matrix saturated with an interstitial fluid \cite{budday20,greiner21}; it is structurally anisotropic \cite{budday20} and it exhibits isotropic, time-dependent mechanical behaviour \cite{rashid14,rashid13,rashid12(2)}. The latter is one of the most pronounced features of brain tissue, manifesting itself primarily through so-called viscoelastic effects. For instance, when brain tissue is deformed rapidly and then held in position, the corresponding stress decreases with time \cite{budday17}, known as stress relaxation. Conversely, when a load is quickly applied and then maintained, the resulting strain increases over time \cite{kang24}, known as creep. This behaviour is commonly observed in many biological tissues and can be attributed to either viscoelastic or plastic effects \cite{GIVERSO_stress_relax,Ambrosi_Preziosi_stressrelax,GIVERSO_creep}. Other time-dependent mechanical effects exhibited by brain tissue include hysteresis and softening resulting from cyclic loading and unloading \cite{budday17}.

The remarkable growth in computational power and technology since the turn of the millennium has led to increased demand from the clinical and biomedical communities for robust, accurate and efficient \textit{in silico} mechanical models that can capture the behaviour of brain tissue in complex real-world scenarios, such as predicting disease progression \cite{budday20}, surgical planning and training \cite{delingette04} and estimating injury risk for contact \cite{ji22} and equestrian sports \cite{connor19}. The wide range of applications above highlights one of the significant challenges facing the computational mechanics community: realistic predictions of brain tissue's mechanical response require sophisticated constitutive models that capture as much of the underlying physics as possible, yet these models must be simple and tractable enough to enable rapid and reliable estimation of their material parameters through calibration with experimental data. 

Viscoelasticity is a major and active area of interest within the field of soft tissue mechanics. For a comprehensive overview of the subject and review of the classical models, the reader is directed to the detailed articles \cite{wineman09,drapaca07}. The most straightforward constitutive theory that can be used to predict the viscoelastic behaviour of brain tissue is linear viscoelasticity, where the instantaneous stress is determined by convolving the strain history with a time-dependent function that depends on brain tissue's material properties \cite{anand20,christensen03}. In reality, brain tissue's viscoelastic response is markedly non-linear and its stress relaxation curves depend on the strain level. A non-linear viscoelastic constitutive theory is, therefore, essential for accurate predictions. Although the literature is replete with non-linear models \cite{wineman09,drapaca07}, they are generally difficult to employ in real-world biomechanical scenarios and numerically costly vis-\`a-vis model fitting and material parameter estimation. To this end, Fung, in his seminal work \cite{fung93}, proposed a compromise approach now known as quasi-linear viscoelasticity (QLV). The QLV model, which falls under the umbrella of the more general Pipkin--Rogers model \cite{wineman09}, is the simplest extension of the linear viscoelastic theory to finite deformations. In contrast to the Pipkin--Rogers model, QLV is limited to materials whose viscous relaxation rates are independent of the instantaneous local strain \cite{depascalis14} and thereby cannot account for the non-linear phenomenon of strain-dependent relaxation commonly observed in biological soft tissues \cite{shearer20,chatelin10,duenwald09,nasseri02}. Nevertheless, its relative simplicity compared to more general non-linear viscoelastic models has led to its widespread use. The QLV model has been employed to model a myriad of biological soft tissues including the skin \cite{karimi16,flynn13}, liver \cite{macmanus19}, brain  \cite{rashid14,rashid13,rashid12(2),sundaresh22,sundaresh21,macmanus20,hosseini19,sahoo14,chatelin13}, lung \cite{daphalapurkar23}, eye \cite{zhang18}, spinal cord \cite{rycman21,yu20,jannesar16}, prostate gland \cite{helisaz24}, eardrum \cite{motallebzadeh13}, oesophagal tissue \cite{yang06}, heart muscle tissue \cite{huyghe91}, ligaments \cite{criscenti15,abramowitch04,funk00}, tendons \cite{duenwald09,bah20}, cartilage \cite{springhetti18,selyutina15}, arteries \cite{giudici23} and membranes \cite{dadgar21}. As noted by De Pascalis \textit{et al.} \cite{depascalis14}, QLV has been criticised for not always exhibiting ``physically reasonable behaviour.'' In that article, the authors thoroughly reappraised the theory and elucidated that its supposed unphysical behaviour stemmed from different interpretations of Fung's original one-dimensional relationship. The main deficiencies in these anterior studies, as summarised by De Pascalis \textit{et al.} \cite{depascalis15}, were using an incorrect QLV relation (especially for incompressible materials) and employing a stress measure other than the second Piola--Kirchhoff stress, which guarantees objectivity. Using the reappraised QLV model (subsequently referred to as modified quasi-linear viscoelasticity or MQLV), De Pascalis \textit{et al}. studied uniaxial tension \cite{depascalis14} and simple shear \cite{depascalis15}, while Righi and Balbi \cite{righi21} considered torsion. Balbi \textit{et al.} \cite{balbi23,balbi18} extended the model to transversely isotropic materials. In contrast to Fung's QLV model, the MQLV model has yet to be validated with experimental data. To the authors' knowledge, the only relevant example is the paper by De Pascalis \textit{et al.} \cite{depascalis18}, which demonstrated that the MQLV model provided a better fit to the relaxation data from inflation tests on murine bladders compared to Fung's model or linear viscoelasticity. Consequently, MQLV's potential for model fitting and material parameter estimation has yet to be fully exploited.

The development of testing protocols to characterise brain tissue's viscoelastic properties presents numerous challenges \cite{budday20,chatelin10}. For instance, brain tissue's fragile, brittle and tacky nature makes it susceptible to damage during sample preparation and testing \cite{budday20}. Furthermore, brain tissue is highly compliant and ultra-soft, which can lead to significant deformation under the action of its own weight, making it difficult to control sample geometry during extraction and testing \cite{budday20}. Additionally, the forces measured during testing often approach the resolution limits of commercially available testing equipment. All these factors and others---for example, age \cite{macmanus17}, species \cite{macmanus17}, anatomical region \cite{menichetti20}, post-mortem storage time \cite{garo07} and temperature \cite{rashid12(3),rashid13(2)}---may contribute to the variations in the mechanical properties of brain tissue reported in the literature.

Budday \textit{et al.} \cite{budday20} and Chatelin \textit{et al.} \cite{chatelin10} provide extensive reviews of the viscoelastic properties of brain tissue as measured in various studies, drawing on over 50 years of research in brain mechanics. These studies demonstrate that stress relaxation in brain tissue has been tested under various deformation modes, including uniaxial tension \cite{rashid14},  uniaxial compression \cite{rashid12(2)}, simple shear \cite{rashid13} and torsion \cite{arbogast98}. Typically, uniaxial tension and compression tests on brain tissue are performed by glueing the ends of a cylindrical sample to the platens \cite{rashid14,rashid12(2)}, restricting lateral expansion or contraction at the sample's ends. This restriction leads to inhomogeneous deformations \cite{rashid14,rashid12}, which cannot be accurately modelled using the analytical solutions available for such tests \cite{depascalis14}; instead, the equations of motion must be solved numerically, complicating model fitting. By contrast, compression tests with lubricated platens can achieve homogeneous deformation conditions but only up to a strain of approximately $10\%$ \cite{balbi19,rashid14}. Another standard testing protocol for brain tissue that can achieve a stretch of more than $60\%$ is simple shear, which is performed by glueing the opposite faces of a cuboidal sample to the platens and recording the shear and normal forces required to move one platen parallel to the other \cite{destrade15}. 
Additionally, surface tractions must be applied to the slanted faces of the deformed sample to prevent bending \cite{destrade23}. In reality, these tractions are never applied; more practically, the effect of deviation from ideal simple shear conditions on the measured shear and normal forces is minimised by using a thin sample whose width is less than four times its height \cite{destrade15,rashid13,destrade23,Britishstandards12}. Furthermore, accurately quantifying the normal force is currently neither feasible nor practical, as it requires recently developed, custom-designed testing equipment \cite{destrade23,yan21}. Thus, in practice, simple shear generates a single dataset for the shear force, similar to uniaxial tension or compression tests that produce a single dataset for the tensile or compressive force. Alternatively, torsion is a more robust and reliable testing protocol that can be readily implemented for brain tissue using commercial devices known as rheometers. These devices measure the torque and normal force required to twist a cylindrical sample, generating two independent datasets (the appearance of a normal force as a result of twisting is an example of the so-called Poynting effect \cite{rivlin49,poynting1909}). The first study to apply this protocol to brain tissue was carried out by Balbi \textit{et al.} \cite{balbi19}, who showed that the instantaneous elastic response of brain tissue in torsion is well-captured by a hyperelastic Mooney--Rivlin model and estimated the corresponding elastic material parameters (the instantaneous shear modulus and Mooney--Rivlin parameters). Although stress relaxation in torsion has been investigated in animal (porcine and bovine) and human brain tissues \cite{budday20,chatelin10}, anterior studies have focused on measuring only the torque, neglecting the normal force. In addition, torsion was modelled as simple shear but only locally is the torsion deformation that of simple shear \cite{balbi19}. Consequently, the potential of torsion as a robust and reliable testing protocol for determining the viscoelastic properties of brain tissue has yet to be fully realised. We note that Narayan \textit{et al.} \cite{narayan12} devised a similar protocol for asphalt binders in torsion, measuring both torque and normal force during relaxation. However, to the best of the authors' knowledge, an analogous protocol for soft tissues has yet to be developed. 

In this work, we exploit the latent potentials of the torsion protocol and the MQLV model to determine the viscoelastic properties of brain tissue. The remainder of this paper is organised as follows. In Section \ref{sec: materials and methods}, we describe the procedure for preparing the cylindrical brain samples and testing them with the rotational rheometer, with the results of the torsion tests presented in Section \ref{sec: experimental results}. In Section \ref{sec: modelling}, we propose a novel fitting procedure for determining brain tissue's viscoelastic material parameters based on the MQLV model. Following this, we present a finite element implementation of the MQLV model in the open-source software FEniCS, which we use to validate our estimates of the viscoelastic material parameters through numerical simulations of the experiments. We discuss the results and summarise the important features of the paper in Section \ref{sec: conclusion}.


\section{Materials and methods}
\label{sec: materials and methods}

In this section, we briefly describe the procedure for preparing the cylindrical brain samples and testing them with the rotational rheometer.


\subsection{Tissue preparation}
\label{sec: tissue preparation}

All experiments were performed using brains from 6 to 9-month-old, mixed-sex sheep obtained from a local European Union-approved slaughterhouse (Athenry Quality Meats Ltd, Galway, Ireland, Approval Number EC2875). As the animals were not sacrificed specifically for this study, ethical approval was not required from the University of Galway’s Research Ethics Committee. 

\begin{figure}[h!]
    {\centerline{\includegraphics[scale=0.75]{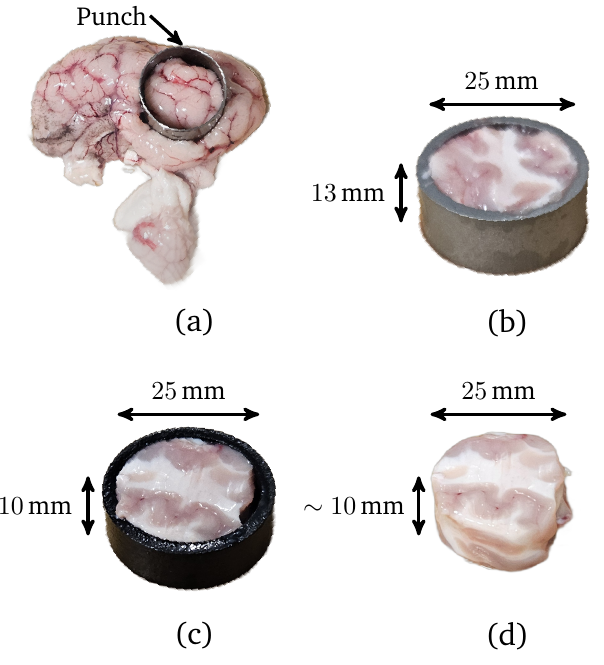}}}
    \caption{Procedure for preparing cylindrical brain samples of radius $12.5\,\,\text{mm}$ and height $10\,\,\text{mm}$ for testing: (a) long cylindrical sample excised from the cerebral hemisphere using a steel punch; (b) top face cut flat using a cutting guide of height $13\,\,\text{mm}$; (c) opposite face cut flat using a cutting guide of height $10\,\,\text{mm}$ and (d) flat cylindrical sample ready for testing.}
    \label{fig: Cutting protocol}
\end{figure}

The brains, which were received as separated cerebral hemispheres, were placed in a phosphate buffered saline (PBS) solution within 1 hour post-mortem to avoid tissue dehydration and maintained at $11$--$15$\,\textdegree \text{C} during transportation. All samples were prepared and tested at room temperature ($19$--$23$\,\textdegree \text{C}). As shown in Fig.~\ref{fig: Cutting protocol}(a), mixed grey and white matter cylindrical samples were excised from the cerebral halves using a sharp $25\,\,\text{mm}$ diameter stainless steel punch. To prepare flat cylindrical samples of radius $R_0=12.5\,\,\text{mm}$ and height $H_0=10\,\,\text{mm}$ for testing, each long sample was first inserted into a cutting guide of height $13\,\,\text{mm}$. The excess brain matter was then removed from the top of the sample using an 8 inch MacroKnife (CellPath, Wales, United Kingdom), as shown in Fig.~\ref{fig: Cutting protocol}(b). Finally, the opposite end of the sample was cut flat with the aid of a cutting guide of height $10\,\,\text{mm}$, as shown in Fig.~\ref{fig: Cutting protocol}(c); the exact heights of the samples were measured before testing. After this, the prepared sample and cerebral hemisphere were placed in a PBS solution. Instead of preparing all the samples at once, each sample was tested immediately after preparation, and then, if possible, another was extracted from the cerebral hemisphere \cite{rashid13}. All samples were tested within 8 hours post-mortem.


\subsection{Mechanical testing}
\label{sec: mechanical testing}

\begin{figure}[h!]
    {\centerline{\includegraphics[scale=0.75]{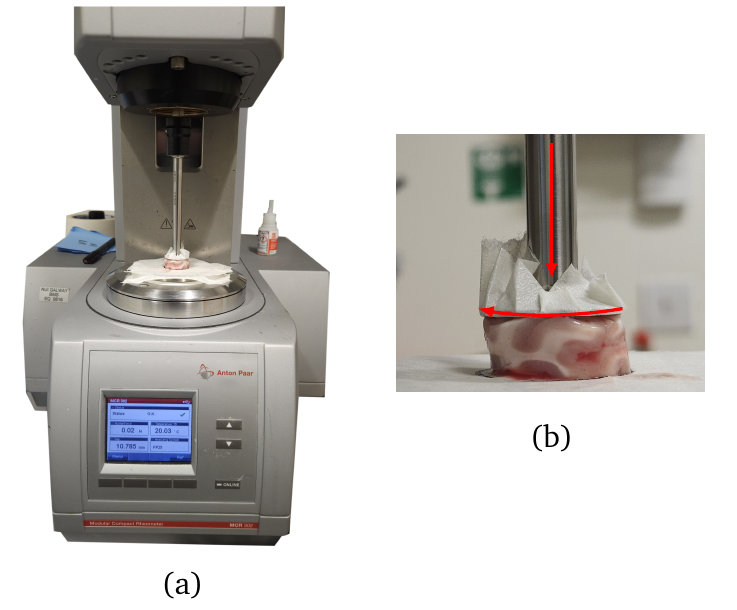}}}
    \caption{(a) Anton Paar MCR 302e rotational rheometer with parallel plate geometry used to perform the torsion tests and (b) side view of a twisted sample during testing.}
    \label{fig: Rheometer}
\end{figure}

An Anton Paar MCR 302e rotational rheometer with parallel plate geometry (Anton Paar, Graz, Austria) was used for the mechanical testing (see Fig.~\ref{fig: Rheometer}). During the tests, the bottom Peltier plate remained fixed, while the motion of the upper plate (which contains the motors and sensors that measure the torque and normal force) was controlled through the software RheoCompass (Version 1.31). To match the dimension of the samples tested, a $25\,\,\text{mm}$ diameter upper plate was used. Masking tape of negligible thickness compared to the sample height was applied to both plates to prevent damage to the rheometer and enable easy removal of the tested samples \cite{balbi19,rashid14,rashid13}. The sample was secured to the tape using a thin layer of cyanoacrylate (RS Radionics, Dublin, Ireland) \cite{rashid14,rashid13,rashid12(2)}. A small pre-compression of approximately $0.03\,\,\text{N}$ was applied by manually lowering the upper plate to ensure proper sample adhesion to the upper and lower plates. A minute and a half was sufficient time for the glue to set, after which the position of the upper plate was slowly adjusted until the normal force read $0\,\,\text{N}$. 

\begin{figure}[h!]
    {\centerline{\includegraphics[scale=0.8]{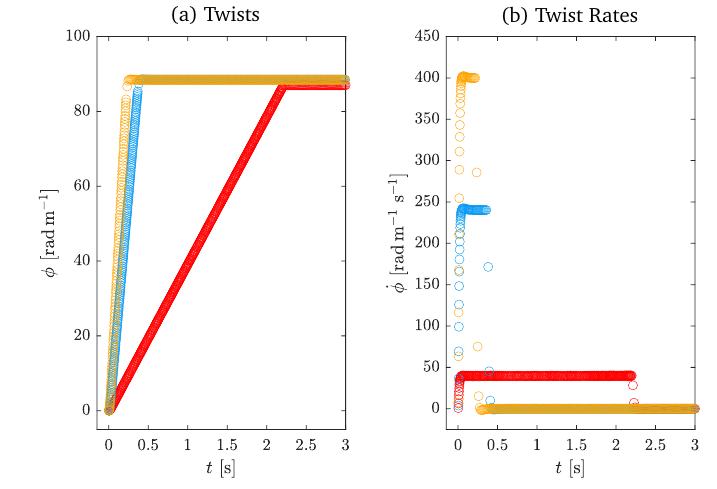}}}
    \caption{(a) Twist and (b) twist rate profiles for our proposed torsion testing protocol. The twist for the red, blue and orange data was increased linearly to a final value of $\phi_0=88\,\,\text{rad}\,\,\text{m}^{-1}$ over durations of $t^{\star} \in \{2.2,0.367,0.22\}\,\,\text{s}$, corresponding to twist rates of $\dot{\phi_0} \in \{40,240,400\}\,\,\text{rad}\,\,\text{m}^{-1}\,\,\text{s}^{-1}$ respectively. After reaching the final twist value, the twist was held constant for $200\,\,\text{s}$.}
    \label{fig: Testing protocol}
\end{figure}

Our torsion testing protocol is summarised in Table \ref{tab: Testing protocol} and illustrated in Fig.~\ref{fig: Testing protocol}. We performed three sets of ramp-and-hold relaxation tests in torsion on the cylindrical samples at varying twist rates of $\dot{\phi_0} \in \{40,240,400\}\,\,\text{rad}\,\,\text{m}^{-1}\,\,\text{s}^{-1}$ (angular velocity of the upper plate per unit deformed height), while keeping the twist fixed at $\phi_0=88\,\,\text{rad}\,\,\text{m}^{-1}$ (angle of rotation per unit deformed height). Each torsion test consisted of two phases: a ramp phase, in which the twist was increased linearly to $\phi_0=88\,\,\text{rad}\,\,\text{m}^{-1}$ over a time $t^{\star} \in \{2.2,0.367,0.22\}\,\,\text{s}$, followed by a hold phase lasting $200\,\,\text{s}$, during which the final value of the twist reached at the end of the ramp phase was maintained. Both the torque $\tau$ and normal force $N_z$ required to twist the sample during the ramp phase and maintain the sample in its deformed state during the hold phase were recorded versus time $t$. No pre-conditioning was performed on the samples, and each was tested only once before being discarded. A total of 30 samples were tested over several campaigns: 10 at $40\,\,\text{rad}\,\,\text{m}^{-1}\,\,\text{s}^{-1}$ (samples $S_1$ to $S_{10}$); 10 at $240\,\,\text{rad}\,\,\text{m}^{-1}\,\,\text{s}^{-1}$ (samples $S_{11}$ to $S_{20}$) and 10 at $400\,\,\text{rad}\,\,\text{m}^{-1}\,\,\text{s}^{-1}$ (samples $S_{21}$ to $S_{30}$).
\begin{table}[h]
\small
  \caption{Torsion testing protocol}
  \begin{tabular*}{\textwidth}{@{\extracolsep{\fill}}l}
    \hline \vspace{0.1cm}
    $\bullet$ \,\,\textbf{Ramp phase:} twist increased linearly to $\phi_0=88\,\,\text{rad}\,\,\text{m}^{-1}$ over durations of $t^{\star} \in \{2.2,0.367,0.22\}\,\,\text{s}$ at twist rates of $\dot{\phi_0} \in \{40,240,400\}\,\,\text{rad}\,\,\text{m}^{-1}\,\,\text{s}^{-1}$  \\[0.1cm]
    $\bullet$ \,\,\textbf{Hold phase:} twist maintained at $\phi_0 = 88\,\,\text{rad}\,\,\text{m}^{-1}$ for $200\,\,\text{s}$  \\ \hline
  \end{tabular*}
  \label{tab: Testing protocol}
\end{table}

\section{Experimental results}
\label{sec: experimental results}

In this section, we present representative torque and normal force data from the rheometer for each of the twist rates $\dot{\phi_0} \in \{40,240,400\}\,\,\text{rad}\,\,\text{m}^{-1}\,\,\text{s}^{-1}$ and describe the filtering procedure applied to the data to prepare it for model fitting and material parameter estimation. 

\begin{figure}[h!]
    {\centerline{\includegraphics[scale=0.8]{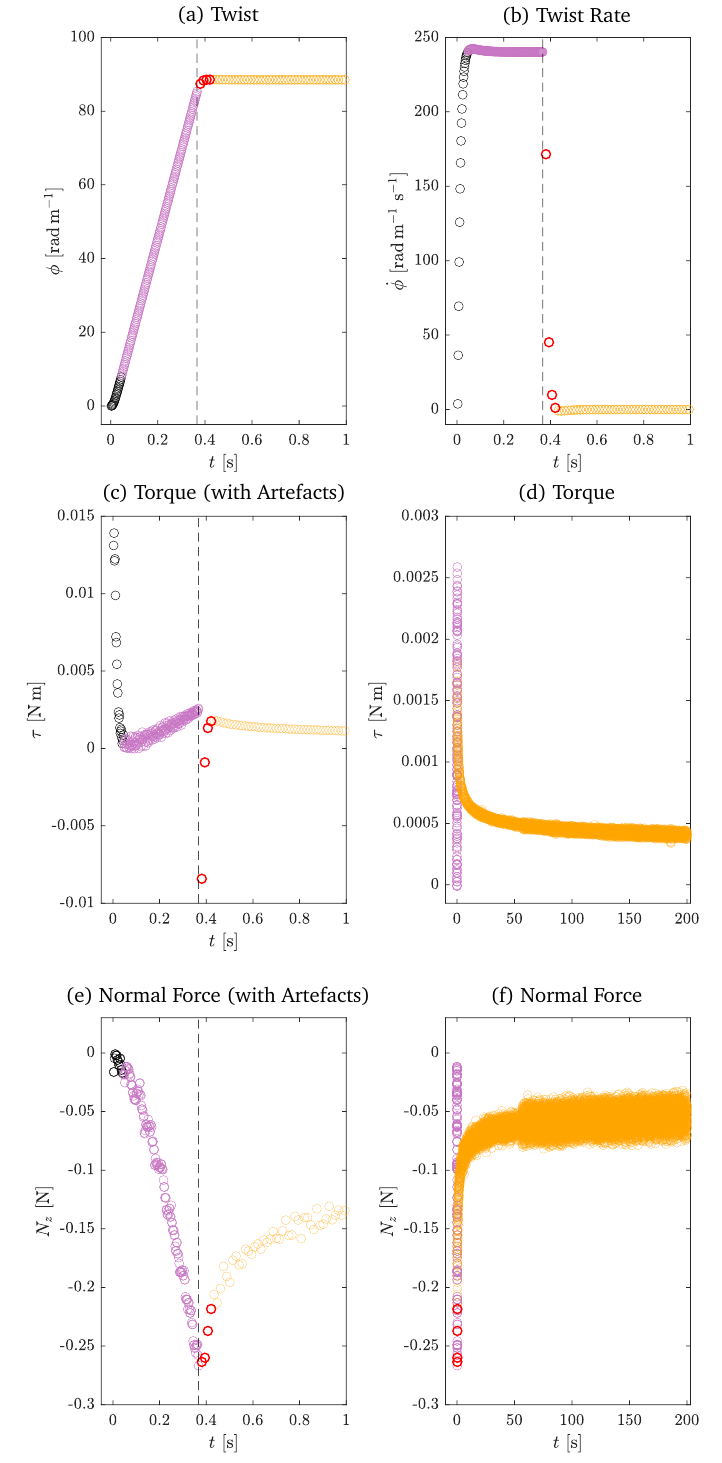}}}
    \caption{Raw output data for sample $S_{16}$ from a torsion test performed at a twist rate of $240\,\,\text{rad}\,\,\text{m}^{-1}\,\,\text{s}^{-1}$: (a,b) twist and twist rate profiles; (c,e) measured torque and normal force for the first second of the test (including experimental artefacts) and (d,f) measured torque and normal force for the entire duration of the test (excluding experimental artefacts). Both the torque data generated when the upper plate was accelerating (black) and decelerating (red) were excluded from the proper torque data in (c), whereas only the normal force data generated when the upper plate was accelerating were excluded from the proper normal force data in (f). A dashed line indicates the end of the ramp phase.}
    \label{fig: Raw output from rheometer}
\end{figure}

As an example, the raw output data for sample $S_{16}$, recorded during a torsion test performed at a twist rate of $\dot{\phi_0} = 240\,\,\text{rad}\,\,\text{m}^{-1}\,\,\text{s}^{-1}$, are presented in Fig.~\ref{fig: Raw output from rheometer}. Fig.~\ref{fig: Raw output from rheometer}(a,b) show the twist $\phi$ and twist rate $\dot{\phi}$ profiles for the test, while Fig.~\ref{fig: Raw output from rheometer}(c,d) display the measured torque $\tau$ and Fig.~\ref{fig: Raw output from rheometer}(e,f) the measured normal force $N_z$. Since the rheometer outputs the normal force exerted by the sample on the upper plate, we changed the sign of the data so that it represents the normal force that must be applied to the sample to maintain the deformation, consistent with the modelling conventions adopted in Section \ref{sec: modelling} and anterior studies \cite{righi21,balbi19}.

From the data presented in Fig.~\ref{fig: Raw output from rheometer}(a,b), we can identify four regions: (i) a region (black data) at the start of the ramp phase, where the upper plate accelerates from rest to the target twist rate of $\dot{\phi_0} = 240\,\,\text{rad}\,\,\text{m}^{-1}\,\,\text{s}^{-1}$; (ii) a region (purple data), where this twist rate is maintained until the twist reaches $\phi_0 = 85\,\,\text{rad}\,\,\text{m}^{-1}$ at the end of the ramp phase at time $t^{\star}=0.367\,\,\text{s}$ (indicated by a dashed line); (iii) a region (red data) at the start of the hold phase, where the upper plate decelerates to rest and (iv) the remainder (orange data) of the hold phase, where the final value of the twist reached at the end of the ramp phase is maintained. We also note experimental artefacts in the torque data in Fig.~\ref{fig: Raw output from rheometer}(c) in regions where there is a rapid change in the twist rate, notably at the start of the ramp phase when the upper plate is accelerating (black data) and the start of the hold phase when it is decelerating (red data); see Fig.~\ref{fig: Raw output from rheometer}(b). Accordingly, these artefacts---potentially due to the inertia of the motors in the upper plate \cite{narayan12}---were excluded from the proper torque data in Fig.~\ref{fig: Raw output from rheometer}(d). Likewise, the proper normal force data in Fig.~\ref{fig: Raw output from rheometer}(c) excludes the less pronounced artefacts in the normal force data at the start of the ramp phase. However, unlike the torque, the normal force data generated at the start of the hold phase does not appear to be adversely affected by the rapid change in twist rate and was therefore included in the proper normal force data in Fig.~\ref{fig: Raw output from rheometer}(c).

During the gamut of tests, the achieved twist rates $\dot{\phi_0}$ were measured as $40.26 \pm 0.42$, $239.95 \pm 0.25$ and $400.06 \pm 0.27\,\,\text{rad}\,\,\text{m}^{-1}\,\,\text{s}^{-1}$ ($\text{mean} \pm \text{SD}$). The corresponding ramp times $t^{\star}$ were set to $2.2$, $0.367$ and $0.22\,\,\text{s}$ to achieve the target twist of $\phi_0=88\,\,\text{rad}\,\,\text{m}^{-1}$ at the end of the ramp phase. However, in practice, the actual twist values reached were slightly lower and decreased with increasing twist rate: $87.48 \pm 0.51$, $85.25 \pm 0.1$ and $83.23 \pm 0.1\,\,\text{rad}\,\,\text{m}^{-1}$. This discrepancy was due to the inertia of the upper plate, which caused the twist to continue increasing slightly at the start of the hold phase while the upper plate decelerated to rest (see Fig.~\ref{fig: Raw output from rheometer}(a)). As a result, the target twist was reached during the hold phase rather than at the end of the ramp phase. This deviation between the target and actual twist values at the end of the ramp phase is a practical limitation of our protocol.  

\begin{figure}[h!]
    {\centerline{\includegraphics[scale=0.80]{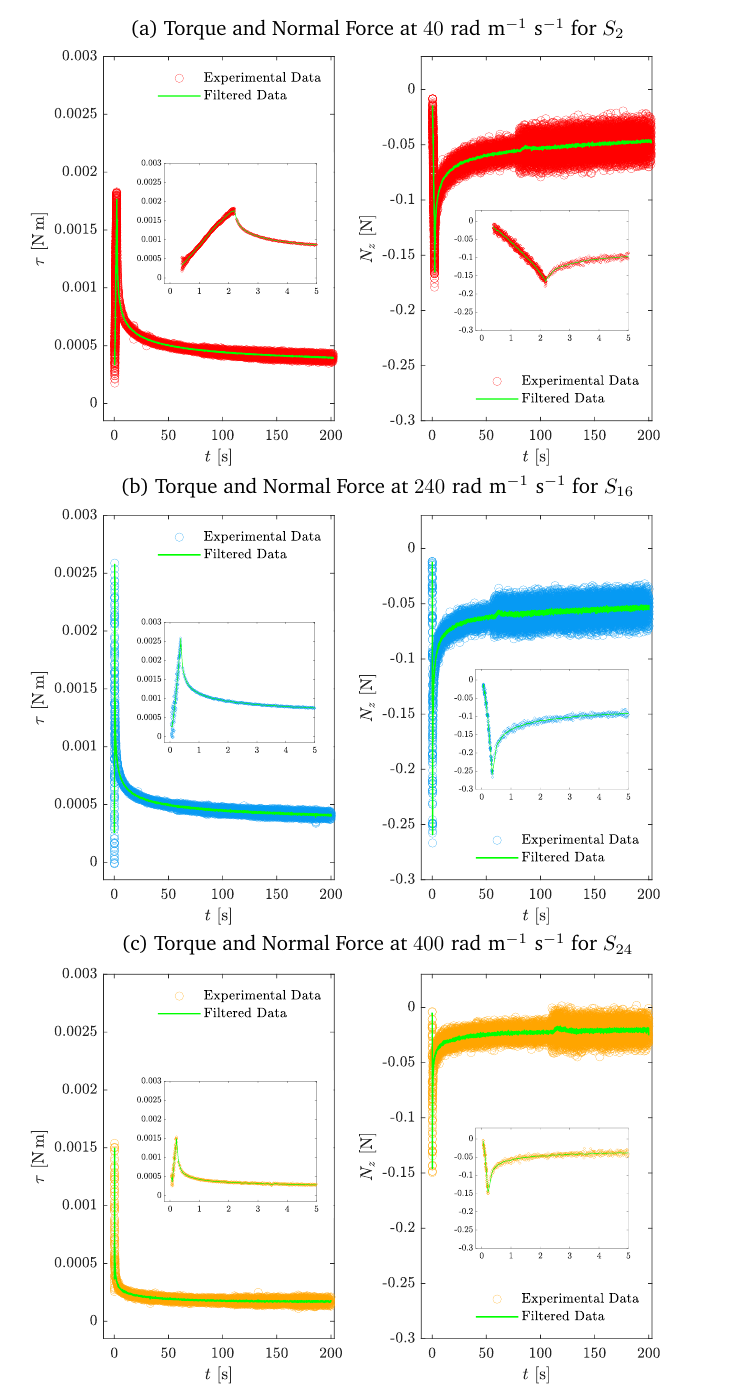}}}
    \caption{Representative torque, normal force and filtered data for samples (a) $S_2$, (b) $S_{16}$ and (c) $S_{24}$ from torsion tests performed at twist rates of $\{40, 240, 400\}\,\,\text{rad}\,\,\text{m}^{-1}\,\,\text{s}^{-1}$. The insets show the ramp phase and the initial part of the hold phase in more detail.}
    \label{fig: Filtered Data}
\end{figure}

Following the protocol of Narayan \textit{et al.} \cite{narayan12}, we estimated the rheometer's torque and normal force resolutions to be approximately $0.15\,\,\text{mN}\,\,\text{m}$ and $0.03\,\,\text{N}$ by performing torsion tests at each of the twist rates $\{40,240,400\}\,\,\text{rad}\,\,\text{m}^{-1}\,\,\text{s}^{-1}$ without any samples between the plates. Another source of noise in the experiments was the attached compressor, which was required for the proper operation of the rheometer. During each test, the compressor would activate to supply fresh compressed air to the air bearings of the rheometer's motors, generating vibrations that increased the noise in the torque and normal force measurements (see Fig.~\ref{fig: Raw output from rheometer}(d,f) at $t \approx 50\,\,\text{s}$). In preparation for model fitting, we smoothed the data by applying a Savitzky--Golay filter using the MATLAB (Version 23.2.0.2485118 (R2023b)) function sgolayfilt \cite{MATLABfilterDoc}. For the $40\,\,\text{rad}\,\,\text{m}^{-1}\,\,\text{s}^{-1}$ data, we used a polynomial order of $5$ and a window length of $61$, whereas for the $240$ and $400\,\,\text{rad}\,\,\text{m}^{-1}\,\,\text{s}^{-1}$ data, we used a polynomial order of $5$ and a window length of $31$. Fig.~\ref{fig: Filtered Data} shows representative torque, normal force and filtered data for samples $S_2$, $S_{16}$ and $S_{24}$ at each twist rate.


\section{Modelling}
\label{sec: modelling}

In this section, we use the MQLV theory to derive analytical expressions for the torque and normal force for a ramp-and-hold test. We then propose a fitting procedure for determining brain tissue's viscoelastic material parameters and apply it to the experimental data. Finally, to validate our fitting results, we perform numerical simulations of the experiments in FEniCS.


\subsection{Theory}
\label{sec: theory}

Here, we calculate the torque and normal force required to maintain an isotropic, incompressible, viscoelastic cylinder in a state of torsion, according to the MQLV theory. Although brain tissue is neither strictly isotropic nor incompressible, experiments indicate these are reasonable assumptions \cite{budday20}. 
 
\begin{figure}[h!]
    {\centerline{\includegraphics[scale=1]{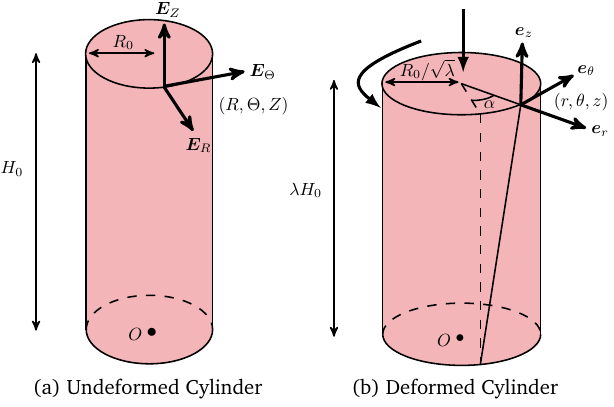}}}
    \caption{(a) Undeformed and (b) deformed cylinder. Torque and normal force must be applied at the end of the cylinder to maintain the torsion deformation.}
    \label{fig: Undeformed and deformed cylinder}
\end{figure}

We consider a cylinder of radius $R_0$ and height $H_0$ subjected to a torsional deformation that takes the point with cylindrical polar coordinates $(R,\Theta,Z)$ in the undeformed configuration (at time $t=0$) to the point with cylindrical polar coordinates $(r,\theta,z)$ in the deformed configuration (at time $t>0$), both relative to a fixed origin $O$. Since the rheometer's normal force sensor has a resolution of approximately $0.03\,\text{N}$, the device cannot detect variations of the normal force within this range. We, therefore, expect the samples to undergo a slight axial contraction before being twisted, even though the upper plate is adjusted until the normal force reads $0\,\text{N}$ before each test \cite{balbi19}. This combined contraction--torsion can be modelled by the following deformation \cite{balbi19,ciarletta14}:   
\begin{equation}
    r(t)=\frac{R}{\sqrt{\lambda}}, \qquad \theta(t)=\Theta+\lambda\phi(t)Z, \qquad z(t)=\lambda Z,  
    \label{Torsion deformation}
\end{equation}
where $0 < \lambda \leq 1$ is the (axial) pre-stretch and the twist $\phi(t)=\alpha(t) \slash \lambda H_0$ is the angle of rotation $\alpha$ per unit deformed height (see Fig.~\ref{fig: Undeformed and deformed cylinder}). The pure torsion case ($\lambda=1$) was considered in \cite{righi21}. By introducing the cylindrical bases $\{\bm{E}_{R},\bm{E}_{\Theta},\bm{E}_{Z}\}$ and $\{\bm{e}_{r},\bm{e}_{\theta},\bm{e}_{z}\}$ for the undeformed and deformed configurations, we can write the deformation gradient $\bm{F}=F_{aA}\,\bm{e}_a\otimes\bm{E}_A$ associated with the deformation \eqref{Torsion deformation} as follows:
\begin{equation}
    \bm{F}(r,t)=\left(\begin{array}{ccc}
                \displaystyle \frac{1}{\sqrt{\lambda}} & 0 & 0 \\[10pt]
                0 & \displaystyle \frac{1}{\sqrt{\lambda}} & r \lambda\phi(t) \\[10pt]
                0 & 0 & \lambda
        \end{array}\right).
        \label{Deformation gradient}
\end{equation}
Various tensors can be computed from the deformation gradient, such as the left Cauchy--Green deformation tensor $\bm{B}=\bm{F}\bm{F}^{\text{T}}$ and right Cauchy--Green deformation tensor $\bm{C}=\bm{F}^{\text{T}}\bm{F}$.

In their experimental study, Balbi \textit{et al}. \cite{balbi19} showed that the instantaneous elastic response of brain tissue in torsion is well-captured by a Mooney--Rivlin strain energy function \cite{depascalis14,mooney40,rivlin48}:
\begin{equation}
    W^{\text{e}}=\frac{\mu_0}{2}\left(\frac{1}{2}+\gamma\right)\left(I_{1}-3\right)+\frac{\mu_0}{2}\left(\frac{1}{2}-\gamma\right)\left(I_{2}-3\right),
    \label{MR strain energy}
\end{equation}
where $\mu_0$ is the instantaneous shear modulus, $\gamma$ is a constant, $I_1=\text{tr}\,\bm{B}$ and $I_2=\text{tr}\,\bm{B}^{-1}$. For this model, the Mooney--Rivlin parameters $c_1$ and $c_2$ are connected to $\mu_0$ and $\gamma$ through $\mu_0=2\left(c_1+c_2\right)$ and $\gamma=1 \slash2-2c_2\slash\mu_0$. The same Mooney--Rivlin behaviour was observed in simple shear \cite{destrade15} and at dynamic strain rates in simple shear \cite{rashid13}, uniaxial tension \cite{rashid14} and uniaxial compression \cite{rashid12(2)}. Under our assumptions, the elastic Cauchy stress corresponding to \eqref{MR strain energy} reads \cite{holzapfel00,ogden97}:
\begin{equation}
    \bm{T}^{\text{e}}=\frac{\mu_0}{2}\left(1+2\gamma\right)\bm{B}-\frac{\mu_0}{2}\left(1-2\gamma\right)\bm{B}^{-1}-p^{\text{e}}\bm{I},
    \label{Elastic stress}
\end{equation}
where $p^{\text{e}}$ is the elastic Lagrange multiplier introduced to enforce the incompressibility constraint ($J=\text{det}\,\bm{F}=1$) and $\bm{I}$ is the second-order identity tensor. 

According to the MQLV theory, the viscoelastic Cauchy stress can be expressed as follows \cite{depascalis14,righi21,depascalis18}:
\begin{equation}
    \bm{T}(r,t)=\bm{F}(r,t)\left(\bm{\Pi}^{\text{e}}_{{\text{D}}}(r,t)+\frac{1}{\mu_0}\int_{0}^{t}\mu^{\,\prime}(t-s)@\bm{\Pi}^{\text{e}}_{{\text{D}}}(r,s)\dd{s}\right)\bm{F}(r,t)^{\text{T}}-p(r,t)\bm{I},   
    \label{Cauchy stress}
\end{equation}
where $p^{\text{e}}$ has been incorporated into the viscoelastic Lagrange multiplier $p$, taken to be a function of $r$ and $t$ only without loss of generality \cite{righi21}. The elastic response in the above is captured by the second Piola--Kirchhoff stress tensor:
\begin{equation}
    \bm{\Pi}^{\text{e}}_{\text{D}}=\bm{F}^{-1}\bm{T}^{\text{e}}_{\text{D}}\bm{F}^{-\text{T}}=\frac{\mu_0}{3}\left(1-2\gamma\right)\bm{C}^{-1}-\frac{\mu_0}{2}\left(1-2\gamma\right)\bm{C}^{-2}+\frac{\mu_0}{2}\left(1+2\gamma\right)\bm{I},
\label{Elastic second Piola-Kirchoff}
\end{equation}
corresponding to the deviatoric Cauchy stress $\bm{T}^{\text{e}}_{\text{D}}=\bm{T}^{\text{e}}-\left(\text{tr}\,\bm{T}^{\text{e}} \slash 3\right)\bm{I}$, while the time-dependent behaviour is associated with the scalar relaxation function $\mu(t)$, taken to be an $n$-term Prony series of the form \cite{rashid13,righi21,morrison23}:
\begin{equation}
    \mu(t)=\mu_{\infty}+\sum_{i=1}^{n}\mu_i\text{e}^{-t \slash \tau_i},
\label{Prony series}
\end{equation}
where $\mu(0)=\mu_0$ and $\mu_{\infty}$ are the instantaneous and long-time shear moduli, $\mu_i$ are the relaxation coefficients and $\tau_i$ are the relaxation times. From \eqref{Deformation gradient}, \eqref{Cauchy stress}, \eqref{Elastic second Piola-Kirchoff} and \eqref{Prony series}, we can determine the components of $\bm{T}$, shown in the Supplementary Material. 

In this work, we assume that the deformation is slow enough that inertial effects can be ignored, although this assumption is not strictly valid during the ramp phase \cite{balbi18,gilchrist13}. In addition, we neglect external body forces. The motion of the cylinder is therefore governed by the momentum balance equation $\text{div}\,\bm{T}=\bm{0}$, where $\text{div}$ is the divergence operator in the deformed configuration \cite{holzapfel00,ogden97}. Upon inspection of the components of the equation of motion, and assuming that the lateral surface of the cylinder is traction-free, we arrive at the following equilibrium problem \cite{righi21}: 
\begin{equation}
\left\{
    \begin{aligned}
    &\pdiff{T_{rr}(r,t)}{r}+\frac{T_{rr}(r,t)-T_{\theta \theta}(r,t)}{r}=0, \\
    &T_{rr}(R_0,t)=0.
    \end{aligned}
    \label{Equilibrium problem}
\right.
\end{equation}
By integrating $\eqref{Equilibrium problem}_1$ subject to $\eqref{Equilibrium problem}_2$, we can determine the Lagrange multiplier $p$ (see the Supplementary Material).

Given the expressions for the stress components $T_{\theta z}$ and $T_{zz}$ in the Supplementary Material, the torque \newline $\tau(t)=2\pi\int_{0}^{R_0 \slash \sqrt{\lambda}}r^2T_{\theta z}(r,t)\dd{r}$ and normal force $N_z(t)=2\pi\int_{0}^{R_0 \slash \sqrt{\lambda}}rT_{zz}(r,t)\dd{r}$ required to maintain the deformation \eqref{Torsion deformation} can be determined by direct integration, with explicit expressions shown in the Supplementary Material. Specialising these equations to a ramp-and-hold test, which corresponds to a twist history of the form (see Fig.~\ref{fig: Testing protocol}(a)):
\begin{equation}
    \phi(t)=\begin{cases}
                \displaystyle  \frac{\phi_{0}t}{t^{\star}}, & 0 \leq t \leq t^{\star} \\
                \displaystyle \phi_{0}, & t>t^{\star},
                \end{cases}
    \label{Ramp-and-hold test twist history}
\end{equation}
we find that the torque and normal force during the hold phase ($t>t^{\star}$) read respectively:
{
\begin{align}
    \tau^{\textrm{hold}}(t)&=\frac{\pi\mu_{\infty}R^4_0}{4}A(\lambda,\gamma)\phi_{0}+\frac{\pi R^4_0}{12\lambda^3t^{\star}}A(\lambda,\gamma)\sum_{i=1}^{n}\mu_{i}\text{e}^{-t \slash \tau_{i}}\big[2(\lambda^3-1)t^{\star}+(\lambda^3+2)\,\tau_{i}\,(\text{e}^{t^{\star} \slash \tau_{i}}-1)\big]\phi_0 \nonumber \\
    &+\frac{\pi R^6_0}{18\lambda t^{\star3}}B(\lambda,\gamma)\sum_{i=1}^{n}\mu_{i}\tau_{i}\text{e}^{-t \slash \tau_{i}}\big[-2\tau_{i}\left(t^{\star}+3\tau_{i}\right)+\text{e}^{t^{\star} \slash \tau_{i}}\,\big(t^{\star 2}-4\tau_{i}t^{\star}+6\tau^{2}_{i}\big) \big]\phi^{3}_{0}
    \label{Hold phase torque} 
\end{align}
}
and
\begin{align}
    N^{\textrm{hold}}_z(t)&=\pi\mu(t)R^2_0\left(\frac{\lambda^3-1}{2\lambda^2}\right)A(\lambda,\gamma)-\frac{\pi\mu_{\infty}R^4_0}{8}B(\lambda,\gamma)\phi^2_{0}-\frac{\pi R^4_0}{12\lambda^4t^{\star 2}}\sum_{i=1}^{n}\mu_{i}\text{e}^{-t \slash \tau_{i}}\big[\left(1-2\gamma\right)\big\{\text{e}^{t^{\star} \slash \tau_{i}}\bigl(3\lambda^3\tau_it^{\star}-C(\lambda)\tau^2_i\bigr) \nonumber \\
    &-D(\lambda)t^{\star 2}+E(\lambda)\tau_it^{\star}+C(\lambda)\tau^2_i\big\}+2\lambda(\lambda^3+2)\tau^2_i\text{e}^{t^{\star} \slash \tau_{i}}+2\lambda(\lambda^3-1)t^{\star 2}-2\lambda(\lambda^3+2)\tau_it^{\star}-2\lambda(\lambda^3+2)\tau^2_i\big]\phi^2_{0} \nonumber \\
    &-\frac{\pi R^6_0}{18\lambda t^{\star 4}}B(\lambda,\gamma)\sum_{i=1}^{n}\mu_{i}\tau^2_i\text{e}^{-t \slash \tau_{i}}\bigl[\text{e}^{t^{\star} \slash \tau_{i}}(t^{\star 2}-6\tau_it^{\star}+12\tau^2_i)-t^{\star 2}-6\tau_it^{\star}-12\tau^2_i\bigr]\phi^4_{0},
    \label{Hold phase normal force}
\end{align}
where the expressions for the functions $A,B,C,D$ and $E$ are given in Appendix \hyperref[sec: appendix a]{A}. Analogous expressions for the ramp phase ($0 \leq t \leq t^{\star}$) are shown in the Supplementary Material. The results of Righi and Balbi \cite{righi21} are recovered by setting $\lambda=1$.


\subsection{Material parameter estimation}
\label{sec: material parameter estimation}

Righi and Balbi \cite{righi21} proposed a fitting procedure for estimating brain tissue's viscoelastic properties in torsion based on the MQLV model for the case when there is no pre-stretch ($\lambda=1$). In their method, the material parameters $\mu_{\infty}$ and $c_2$ are determined from the long-time (asymptotic) values of the torque and normal force, while $\mu_i$ and $\tau_i$ are obtained by fitting the measured torque for the hold phase to the MQLV analytical prediction. Since this procedure does not incorporate a pre-stretch, an alternative fitting procedure that accounts for the effect of pre-stretch on the torque and normal force is required to fit the experimental data accurately.

To this end, we estimated the complete set of viscoelastic material parameters by simultaneously fitting the torque and normal force datasets for each of the twist rates $\dot{\phi_0} \in \{40,240,400\}\,\,\text{rad}\,\,\text{m}^{-1}\,\,\text{s}^{-1}$ to the MQLV analytical predictions \eqref{Hold phase torque} and \eqref{Hold phase normal force}. This was done using the MATLAB function fmincon \cite{MATLABfminconDoc}, which implements an algorithm based on the interior-point method for solving constrained non-linear optimisation problems. Due to the increased noise during the ramp phase, we confined the fitting to the final ramp phase point and the entire hold phase, i.e. $t \geq t^{\star}$. For the twists $\phi_0$, we used the measured values at the end of the ramp phase rather than the target value. We minimised the following objective function:  
\begin{equation}
    \chi^{2}=\sum_{i=1}^{n_1}\left(\frac{\tau^{\text{MQLV}}_{i}-\tau^{\text{Exp}}_{i}}{\tau^{\text{Exp}}_{i}}\right)^2+\sum_{i=1}^{n_2}\left(\frac{N^{\text{MQLV}}_{z,\,i}-N^{\text{Exp}}_{z,\,i}}{N^{\text{Exp}}_{z,\,i}}\right)^2,
    \label{Objective function}
\end{equation}
where $n_1$ and $n_2$ are the numbers of considered torque and normal force datapoints $\tau^{\text{Exp}}_{i}$ and $N^{\text{Exp}}_{z,\,i}$, while  $\tau^{\text{MQLV}}_{i}$ and $N^{\text{MQLV}}_{z,\,i}$ are the corresponding MQLV analytical predictions. This method is similar to that used by Anssari-Benam \textit{et al.} \cite{anssari22}. The advantage of this approach, which minimises the relative error, over the more common approach of minimising the absolute error was discussed by Destrade \textit{et al.} \cite{destrade17}. We estimated the Prony parameters ($\mu_{\infty}$, $\mu_i$ and $\tau_i$) from a $4$-term Prony series. To ensure that the fitting results were physically plausible, we constrained the Prony and Mooney--Rivlin parameters to be positive and limited the pre-stretch to no more than $1\%$. 

The fitting results for each of the twist rates are shown in Tables \ref{tab: Sample parameters for 40 twist rate}, \ref{tab: Sample parameters for 240 twist rate} and \ref{tab: Sample parameters for 400 twist rate}. For the parameters $\mu_{0}$, $\mu_{\infty}$, $\mu_i$ and $c_2$, we report the mean and standard deviation. For the remaining parameters $\tau_i$ and $c_1$, which have a non-symmetric distribution due to potential outliers, we report the median and interquartile range (IQR). The goodness of fit was assessed based on the relative errors ($\%$) in the torque and normal force, defined by $\text{err}_{\tau}=\abs{\left(\tau^{\text{MQLV}}-\tau^{\text{Exp}}\right) \slash \tau^{\text{Exp}}}\times 100$ and $\text{err}_{N_{z}}=\abs{\left(N^{\text{MQLV}}_{z}-N^{\text{Exp}}_{z}\right) \slash N^{\text{Exp}}_{z}}\times 100$. The MQLV model provides good fits to both the torque and normal force datasets simultaneously, exhibiting small to moderate maximal relative errors over the fitting range for all samples, with the exception of $S_{24}$ (see Tables \ref{tab: Sample parameters for 40 twist rate}, \ref{tab: Sample parameters for 240 twist rate} and \ref{tab: Sample parameters for 400 twist rate}). We note that the relative errors observed here for the MQLV model are similar to those reported by Anssari-Benam \textit{et al.} \cite{anssari22}, who simultaneously fit uniaxial tension/compression and simple shear datasets for brain tissue to different hyperelastic models.    

\begin{figure}[h!]
    {\centerline{\includegraphics[scale=0.7]{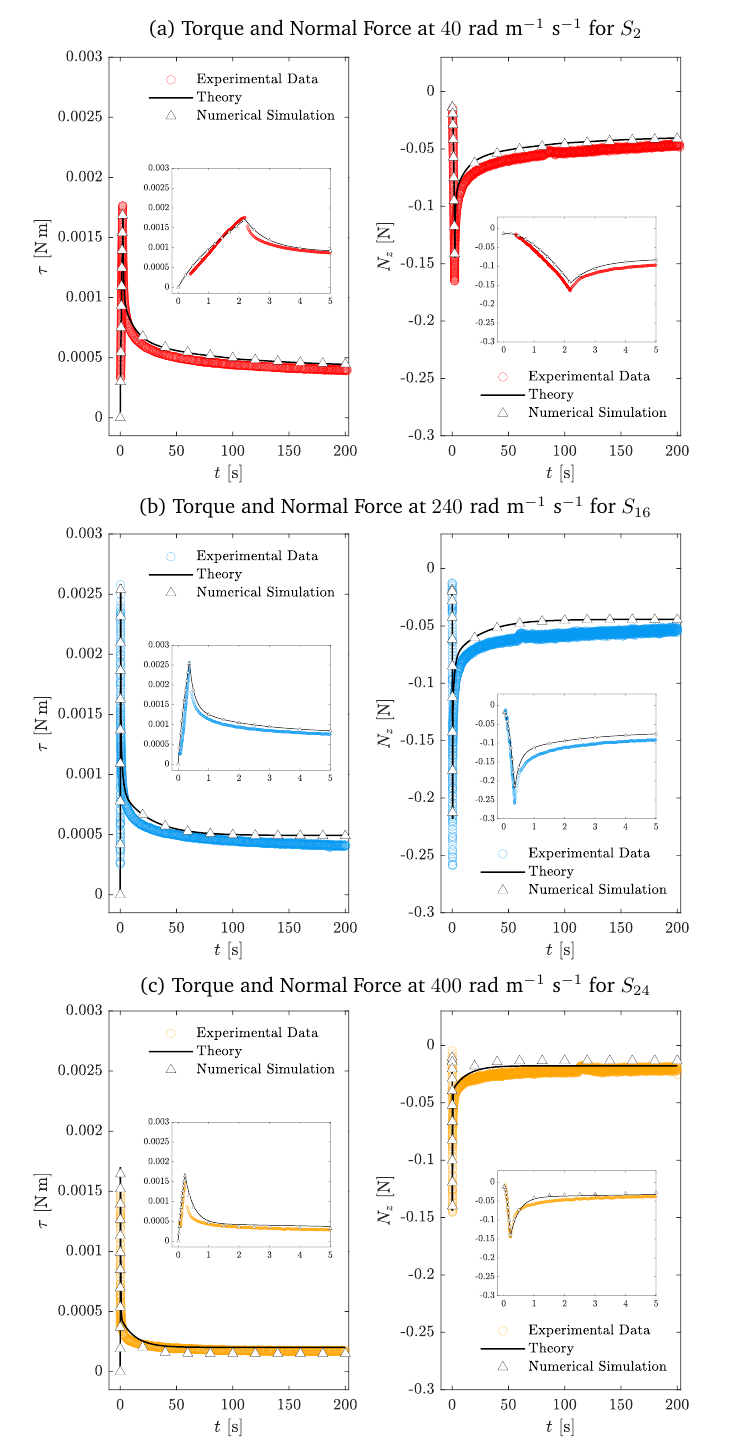}}}
    \caption{Comparison of the resultant torque $\tau$ and normal force $N_z$ for samples (a) $S_2$, (b) $S_{16}$ and (c) $S_{24}$ at twist rates of $\{40, 240, 400\}\,\,\text{rad}\,\,\text{m}^{-1}\,\,\text{s}^{-1}$. Experimental data are denoted by circles, analytical predictions using the MQLV model by solid black lines and the results of the numerical simulations in FEniCS by triangles. The insets show the ramp phase and the initial part of the hold phase in more detail.}
    \label{fig: Comparison of results}
\end{figure}

\begin{figure}[h!]
    {\centerline{\includegraphics[scale=0.7]{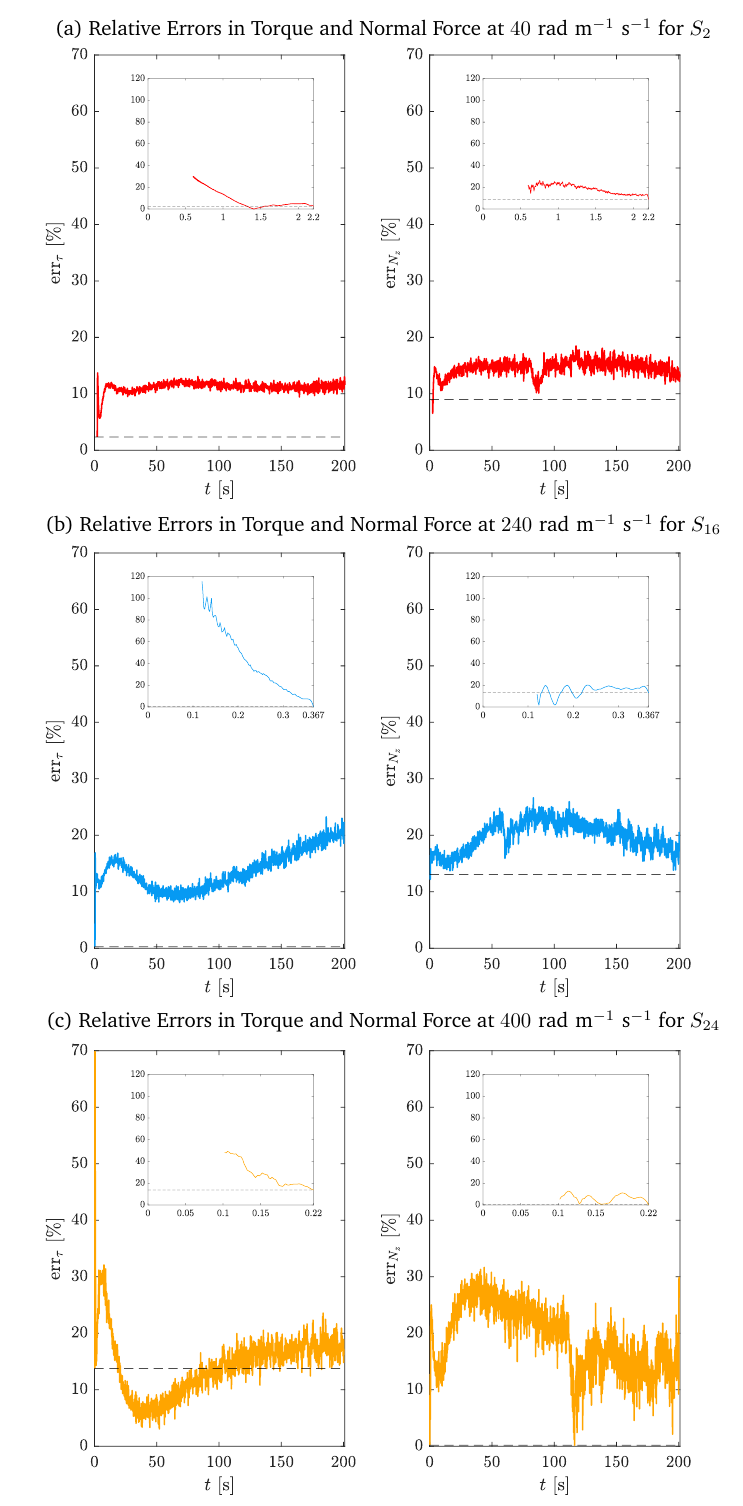}}}
    \caption{Relative errors in the torque $\text{err}_{\tau}$ and force $\text{err}_{N_{z}}$ for samples (a) $S_2$, (b) $S_{16}$ and (c) $S_{24}$ at twist rates of $\{40, 240, 400\}\,\,\text{rad}\,\,\text{m}^{-1}\,\,\text{s}^{-1}$. The insets show the ramp phase in more detail. The dashed lines indicate the relative errors at the end of the ramp phase.}
    \label{fig: Errors}
\end{figure}

As an example, Fig.~\ref{fig: Comparison of results} shows the MQLV analytical predictions for the torque and normal force for samples $S_2$, $S_{16}$ and $S_{24}$, while Fig.~\ref{fig: Errors} displays the corresponding relative errors. As the insets in Fig.~\ref{fig: Errors} indicate, the relative errors in the torque are initially moderate to large but rapidly decrease towards the end of the ramp phase. By contrast, the relative errors in the force during the ramp phase are small to moderate. Over the fitting range, comprised of the final ramp phase point and the entire hold phase, the relative errors in both the torque and normal force are similarly small to moderate. The exception is sample $S_{24}$, where the relative error in the torque at the start of the hold phase increases sharply before decreasing at a similar rate. This is likely due to the high twist rate of $400\,\,\text{rad}\,\,\text{m}^{-1}\,\,\text{s}^{-1}$ and deviations from the ideal cylindrical geometry assumed by the rheometer and MQLV model. Nevertheless, the relative errors for the remainder of the hold phase are in line with those of samples $S_2$ and $S_{16}$.

\begin{table*}[h!]
\tiny
  \caption{Heights, estimated MQLV parameters and maximal relative errors in the torque and normal force over the fitting range for samples tested at $40\,\,\text{rad}\,\,\text{m}^{-1}\,\,\text{s}^{-1}$}
  {\fontsize{3.5}{2}
 \begin{tabular*}{\textwidth}{@{\extracolsep{\fill}}ccccccccccccccccc}
    \hline
    $\text{Sample}$ & $H_0$ $[\text{mm}]$ & $\lambda$ & $\mu_{0}$ $[\text{Pa}]$ & $\mu_{\infty}$ $[\text{Pa}]$ & $\mu_{1}$ $[\text{Pa}]$ & $\mu_{2}$ $[\text{Pa}]$ &$\mu_{3}$ $[\text{Pa}]$ & $\mu_{4}$ $[\text{Pa}]$ & $\tau_{1}$ $[\text{s}]$ & $\tau_{2}$ $[\text{s}]$ &$\tau_{3}$ $[\text{s}]$ & $\tau_{4}$ $[\text{s}]$ & $c_1$ $[\mu\text{Pa}]$ & $c_2$ $[\text{Pa}]$ & $\max\limits_{\,\,t\geq t^{\star}}\text{err}_{\tau}$ $[\%]$ & $\max\limits_{\,\,t\geq t^{\star}}\text{err}_{N_{z}}$ $[\%]$ \\
    \hline 
    $S_{1}$ & $9.4065$ & $0.99$ & $651.16$ & $105.67$ & $213.34$ & $192.34$ & $85.831$ & $53.892$ & $0.8397$ & $0.244$ & $6.4136$ & $55.312$ & $0.10225$ & $325.58$ & $11.609$ & $16.054$  \\ 
    $S_{2}$ & $8.8319$ & $0.99$ & $917.69$ & $125.55$ & $606.76$ & $111.28$ & $73.07$ & $1.0213$ & $0.6811$ & $9.5764$ & $82.406$ & $48.45$ & $2.0793$ & $458.84$ & $13.725$ & $18.992$  \\ 
    $S_{3}$ & $8.7603$ & $0.99$ & $973.9$ & $128.37$ & $583.34$ & $95.025$ & $91.207$ & $75.954$ & $0.443$ & $9.4962$ & $2.3379$ & $75.888$ & $0.4711$ & $486.95$ & $13.415$ & $18.628$  \\ 
    $S_{4}$ & $9.2659$ & $0.99$ & $1048$ & $142.97$ & $628.57$ & $110.42$ & $91.321$ & $74.705$ & $0.4005$ & $2.398$ & $10.442$ & $72.484$ & $0.13183$ & $523.99$ & $14.281$ & $19.4$  \\ 
    $S_{5}$ & $8.9006$ & $0.99$ & $1308.8$ & $161.91$ & $455.32$ & $455.31$ & $137.22$ & $99.052$ & $0.6455$ & $0.1715$ & $5.9739$ & $58.745$ & $0.021216$ & $654.41$ & $18.485$ & $25.642$  \\ 
    $S_{6}$ & $9.3734$ & $0.99$ & $581.02$ & $64.313$ & $368.49$ & $66.885$ & $44.183$ & $37.148$ & $0.3159$ & $1.782$ & $8.4904$ & $65.141$ & $1.5548$ & $290.51$ & $10.915$ & $17.506$  \\ 
    $S_7$ & $8.3203$ & $0.99$ & $1195.4$ & $116.2$ & $816.85$ & $137.58$ & $70.991$ & $53.785$ & $0.3639$ & $2.5286$ & $10.9$ & $63.17$ & $0.028749$ & $597.7$ & $14.239$ & $19.997$  \\ 
    $S_{8}$ & $8.4103$ & $0.99$ & $739.23$ & $107.05$ & $258.68$ & $242.45$ & $76.34$ & $54.709$ & $0.194$ & $0.7838$ & $6.3087$ & $43.982$ & $0.003838$ & $369.61$  & $15.575$ & $28.523$  \\  
    $S_{9}$ & $7.9813$ & $0.99$ & $1216$ & $149.53$ & $704.25$ & $187.15$ & $103.02$ & $72.031$ & $0.3057$ & $1.9388$ & $10.93$ & $78.995$ & $0.0094199$ & $607.99$ & $15.803$ & $21.589$ \\
    $S_{10}$ & $8.3523$ & $0.99$ & $754.36$ & $71.934$ & $437.68$ & $151.71$ & $55.116$ & $37.911$ & $0.2027$ & $1.1376$ & $8.1273$ & $58.425$ & $4.2268$ & $377.18$  & $12.582$ & $21.687$ \\
    $\text{Mean}\pm\text{SD}\, \slash\, \text{Median}\,\,(\text{IQR})$ & --- & --- & $943 \pm 258$ & $117 \pm 32$ & $496 \pm 177$ & $188 \pm 114$ & $85 \pm 26$ & $57 \pm 27$ & $0.36\,\,(0.25$--$0.59)$ &  $1.52\,\,(0.87$--$2.28)$ &  $8.15\,\,(6.33$--$9.95)$ &  $58.59\,\,(55.42$--$70.65)$ &  $0.3\,\,(0.041$--$1.36)$ & $471 \pm 129$ & $14.06 \pm 2.2$ & $20.8 \pm 3.78$ \\
    \hline
  \end{tabular*}
  }
  \label{tab: Sample parameters for 40 twist rate}
\end{table*}

\begin{table*}[h!]
\tiny
  \caption{Heights, estimated MQLV parameters and maximal relative errors in the torque and normal force over the fitting range for samples tested at $240\,\,\text{rad}\,\,\text{m}^{-1}\,\,\text{s}^{-1}$}
  {\fontsize{3.5}{2}
  \begin{tabular*}{\textwidth}{@{\extracolsep{\fill}}ccccccccccccccccc}
    \hline
    $\text{Sample}$ & $H_0$ $[\text{mm}]$ & $\lambda$ & $\mu_{0}$ $[\text{Pa}]$ & $\mu_{\infty}$ $[\text{Pa}]$ & $\mu_{1}$ $[\text{Pa}]$ & $\mu_{2}$ $[\text{Pa}]$ &$\mu_{3}$ $[\text{Pa}]$ & $\mu_{4}$ $[\text{Pa}]$ & $\tau_{1}$ $[\text{s}]$ & $\tau_{2}$ $[\text{s}]$ &$\tau_{3}$ $[\text{s}]$ & $\tau_{4}$ $[\text{s}]$ & $c_1$ $[\mu\text{Pa}]$ & $c_2$ $[\text{Pa}]$ & $\max\limits_{\,\,t\geq t^{\star}}\text{err}_{\tau}$ $[\%]$ & $\max\limits_{\,\,t\geq t^{\star}}\text{err}_{N_{z}}$ $[\%]$ \\
    \hline 
    $S_{11}$ & $8.1853$ & $0.99$ & $1020$ & $88.511$ & $359.74$ & $359.41$ & $132.2$ & $80.179$ & $0.1117$ & $0.0941$ & $1.2096$ & $20.497$ & $91.948$ & $510.02$ & $19.497$  & $18.886$ \\ 
    $S_{12}$ & $8.7724$ & $0.99$ & $861.35$ & $105.47$ & $334.31$ & $315.94$ & $61.799$ & $43.828$ & $0.12$ & $0.5626$ & $17.937$ & $19.834$ & $1.39$ & $430.68$ & $30.806$ & $31.609$  \\ 
    $S_{13}$ & $8.8588$ &  $0.99$ & $1253.8$ & $177.06$ & $740.79$ & $194.66$ & $141.29$ & $3.9931 \times 10^{-4}$ & $0.209$ & $2.9542$ & $38.93$ & $18.451$ & $10.197$ & $626.9$ & $18.992$  & $22.226$   \\ 
    $S_{14}$ & $9.9285$ & $0.99$ & $714.17$ & $107.18$ & $181.16$ & $154.9$ & $153.56$ & $117.37$ & $0.4112$ & $0.4929$ & $0.259$ & $15.306$ & $28.933$ & $357.08$ & $35.938$ & $31.804$  \\ 
    $S_{15}$ & $9.7461$ & $0.99$ & $809.24$ & $110.98$ & $267.8$ & $240.32$ & $108.64$ & $81.494$ & $0.2705$ & $0.1037$ & $2.187$ & $28.974$ & $6.2476$ & $335.51$ & $23.35$ & $30.089$  \\ 
    $S_{16}$ & $9.4673$ & $0.99$ & $1264.4$ & $148.86$ & $431.27$ & $399.57$ & $170.89$ & $113.81$ & $0.1179$ & $0.2148$ & $1.7763$ & $25.905$ & $11.289$ & $632.2$ & $22.962$ & $26.81$  \\ 
    $S_{17}$ & $9.4488$ & $0.99$ & $941.46$ & $142.06$ & $605.37$ & $130.48$ & $63.541$ & $0.0062$ & $0.2915$ & $7.3912$ & $89.376$ & $105.57$ & $1.3055$ & $470.73$  & $28.384$ & $34.375$  \\ 
    $S_{18}$ &  $11.341$ & $0.99$ & $1159.5$ & $134.3$ & $384.24$ & $381.77$ & $161.17$ & $197.97$ & $0.1539$ & $0.1917$ & $2.0081$ & $33.052$ & $8.7216$ & $579.73$ & $25.468$ & $31.333$  \\  
    $S_{19}$ & $8.1197$ & $0.99$ & $1216$ & $149.53$ & $704.25$ & $187.15$ & $103.02$ & $72.031$ & $0.3057$ & $1.9388$ & $10.93$ & $78.995$ & $0.0094199$ & $607.99$ & $22.622$ & $26.891$ \\
    $S_{20}$ & $9.0278$  & $0.99$ & $1238.3$ & $168.98$ & $389.22$ & $377.06$ & $171.6$ & $131.48$ & $0.3024$ & $0.1172$ & $2.806$ & $36.611$ & $2.1066$ & $619.17$ & $21.601$ & $25.781$ \\
    $\text{Mean}\pm\text{SD}\, \slash\, \text{Median}\,\,(\text{IQR})$ & ---  & --- & $1041 \pm 201$ & $130 \pm 29$ & $447 \pm 198$ & $270 \pm 108$ & $126 \pm 41$ & $77 \pm 66$ & $0.19\,\,(0.13$--$0.27)$ & $0.35\,\,(0.14$--$1.9)$ & $2.5\,\,(1.83$--$29.9)$ & $25.98\,\,(20$--$32.03)$ & $9.45\,\,(3.14$--$11.02)$ & $513 \pm 111$ & $24.96 \pm 5.33$ & $27.98 \pm 4.8$ \\
    \hline
  \end{tabular*}
  }
  \label{tab: Sample parameters for 240 twist rate}
\end{table*}

\begin{table*}[h!]
\tiny
  \caption{Heights, estimated MQLV parameters and maximal relative errors in the torque and normal force over the fitting range for samples tested at $400\,\,\text{rad}\,\,\text{m}^{-1}\,\,\text{s}^{-1}$}
  {\fontsize{3.5}{2}
   \begin{tabular*}{\textwidth}{@{\extracolsep{\fill}}ccccccccccccccccc}
    \hline
    $\text{Sample}$ & $H_0$ $[\text{mm}]$ & $\lambda$ & $\mu_{0}$ $[\text{Pa}]$ & $\mu_{\infty}$ $[\text{Pa}]$ & $\mu_{1}$ $[\text{Pa}]$ & $\mu_{2}$ $[\text{Pa}]$ &$\mu_{3}$ $[\text{Pa}]$ & $\mu_{4}$ $[\text{Pa}]$ & $\tau_{1}$ $[\text{s}]$ & $\tau_{2}$ $[\text{s}]$ &$\tau_{3}$ $[\text{s}]$ & $\tau_{4}$ $[\text{s}]$ & $c_1$ $[\mu\text{Pa}]$ & $c_2$ $[\text{Pa}]$ & $\max\limits_{\,\,t\geq t^{\star}}\text{err}_{\tau}$ $[\%]$ & $\max\limits_{\,\,t\geq t^{\star}}\text{err}_{N_{z}}$ $[\%]$ \\
    \hline 
    $S_{21}$ & $11.402$ & $0.99$ & $989.56$ & $116.08$ & $359.18$ & $344.12$ & $170.18$ & $2.3359 \times 10^{-4}$ & $0.114$ & $0.2938$ & $13.187$ & $12.323$ & $10.829$ & $494.78$ & $29.252$ & $29.349$   \\ 
    $S_{22}$ & $11.931$ & $0.99$ & $774.72$ & $89.185$ & $497.42$ & $113.66$ & $74.464$ & $2.1 \times 10^{-4}$ & $0.185$ & $2.7482$ & $33.279$ & $8.6742$ & $0.92322$ & $387.36$ & $31.812$ & $40.139$  \\ 
    $S_{23}$ & $8.5067$ & $0.99$ & $388.24$ & $58.62$ & $133.86$ & $123.86$ & $71.964$ & $4.4007 \times 10^{-5}$ & $0.6331$ & $6.4661$ & $40.349$ & $60.757$ & $0.0059017$ & $223.98$  & $27.018$ & $34.238$  \\ 
    $S_{24}$ & $10.213$ & $0.99$ & $920.46$ & $62.668$ & $535.86$ & $244.75$ & $77.299$ & $0.0013$ & $0.2815$ & $0.0011$ & $12.829$ & $0.9428$ & $42.44$ & $460.23$  & $69.909$  & $32.426$  \\ 
    $S_{25}$ & $10.58$ & $0.99$ & $1379.5$ & $127.36$ & $370.45$ & $352.02$ & $351.55$ & $178.14$ & $0.0674$ & $0.3406$ & $0.0671$ & $12.155$ & $0.10553$ & $689.77$ & $30.786$ & $36.412$ \\ 
    $S_{26}$ & $11.244$ & $0.99$ & $497.15$ & $65.435$ & $114.21$ & $114.15$ & $112.31$ & $91.044$ & $0.1048$ & $0.4212$ & $0.397$ & $11.621$ & $4.0076$ & $248.57$ & $34.013$ & $37.783$  \\ 
    $S_{27}$ & $8.1004$ & $0.99$ & $1005.8$ & $83.491$ & $264.95$ & $260.9$ & $256.02$ & $140.45$ & $0.1004$ & $0.0621$ & $0.317$ & $11.814$ & $6.3762$ & $502.91$ & $35.692$ & $44.153$ \\
    $S_{28}$ & $8.6911$ & $0.99$ & $1178.7$ & $157.1$ & $304.35$ & $274.33$ & $267.27$ & $175.67$ & $0.4461$ & $0.0826$ & $0.0828$ & $14.472$ & $9.6969$ & $589.36$ & $28.884$ & $30.932$  \\
    $S_{29}$ & $8.8104$ & $0.99$ & $1495.9$ & $137.88$ & $512.7$ & $507.84$ & $216.32$ & $121.2$ & $0.0826$ & $0.1105$ & $1.2196$ & $24.011$ & $3.8003$ & $747.97$ & $19.828$ & $19.538$ \\
    $S_{30}$ & $8.8442$ & $0.99$ & $1161.9$ & $125$ & $375.42$ & $374.91$ & $170.68$ & $115.85$ & $0.114$ & $0.1037$ & $0.9253$ & $19.031$ & $7.6129$ & $580.93$ & $30.257$ & $37.282$ \\
    $\text{Mean}\pm\text{SD}\, \slash\, \text{Median}\,\,(\text{IQR})$ & ---  & --- & $979 \pm 354$ & $102 \pm 35$ & $347 \pm 147$ & $271 \pm 129$ & $177 \pm 96$ & $82 \pm 75$ & $0.11\,\,(0.1$--$0.26)$ & $0.2\,\,(0.088$--$0.0.4)$& $1.07\,\,(0.34$--$13.1)$ & $12.24\,\,(11.67$--$17.89)$ & $5.19\,\,(1.64$--$9.18)$ & $493 \pm 172$ & $33.75 \pm 13.41$ & $34.23 \pm 6.78$ \\
    \hline
  \end{tabular*}
  }
  \label{tab: Sample parameters for 400 twist rate}
\end{table*}

\subsection{Computational validation}
\label{sec: computational validation}  

To validate the fitting results from Section \ref{sec: material parameter estimation}, we conducted brain torsion simulations using the open-source software FEniCS \cite{alnaes15, logg12}. FEniCS provides a high-level Python and C\texttt{++} interface for solving partial differential equations via the finite element method. For the numerical implementation of the problem, the governing equations for the torsion of a solid cylinder were expressed in Cartesian coordinates and formulated using a Lagrangian description of motion, where all quantities of interest are represented in material coordinates.

In this framework, the governing equations are written as $\text{Div}\,\bm{P} = \bm{0}$ on the undeformed cylinder, where $\text{Div}$ represents the divergence operator with respect to the undeformed configuration and $\bm{P} = J \bm{T} \bm{F}^{-\text{T}}$ denotes the first Piola--Kirchhoff stress tensor \cite{holzapfel00,ogden97}. To enforce incompressibility, we included the constraint $J = 1$ in the model. Before applying torsion, an initial step was introduced to simulate pre-compression, involving an axial pre-stretch of $\lambda = 0.99$. For the torsion simulation, a reference point at the centre of the cylinder's top surface was defined and coupled to all other points on the top surface to ensure uniform rotational displacement about the longitudinal axis. Appropriate boundary conditions were applied as follows: the bottom surface of the cylinder was fixed to prevent displacement, while the lateral surface remained traction-free throughout the simulation.

Numerical simulations were performed for samples $S_2$, $S_{16}$ and $S_{24}$ at twist rates of $\{40, 240, 400\}\,\,\text{rad}\,\,\text{m}^{-1}\,\,\text{s}^{-1}$. The twist was ramped linearly to match experimental values at the end of the ramp phase over time periods of $\{2.2, 0.367, 0.22\}\,\,\text{s}$, after which it was held constant (see Fig.~\ref{fig: Testing protocol}). The initial cylindrical geometry for each sample was defined with a radius of $R_0 = 12.5\,\,\text{mm}$, while the heights $H_0$ were specified in Tables \ref{tab: Sample parameters for 40 twist rate}, \ref{tab: Sample parameters for 240 twist rate} and \ref{tab: Sample parameters for 400 twist rate}. Computational meshes were generated in FEniCS, consisting of $16,809$, $17,391$ and $12,547$ tetrahedral elements for the respective samples. To enable a solution via the finite element method, the weak formulation of the Lagrangian model was derived and a spatial discretisation of the continuous variational problem was introduced. Specifically, quadratic tetrahedral elements $\mathbb{P}_2$ were employed for the displacement field, while linear tetrahedral elements $\mathbb{P}_1$ were used for the pressure field. This combination is commonly referred to as Taylor--Hood elements. The time step was initially set to a small value ($\Delta t \in (0.001, 0.01)\,\,\text{s}$) during the early simulation phases to accurately capture large deformations. As deformation rates decreased, the time step was progressively increased to enhance computational efficiency. The estimated material parameters, determined via the MQLV theory, were assigned as specified in Tables \ref{tab: Sample parameters for 40 twist rate}, \ref{tab: Sample parameters for 240 twist rate} and \ref{tab: Sample parameters for 400 twist rate}. The torque and normal force were then computed with ParaView \cite{Paraview}, an open-source application for interactive scientific visualisation.

\begin{figure}[h!]
    {\centerline{\includegraphics[scale=0.4]{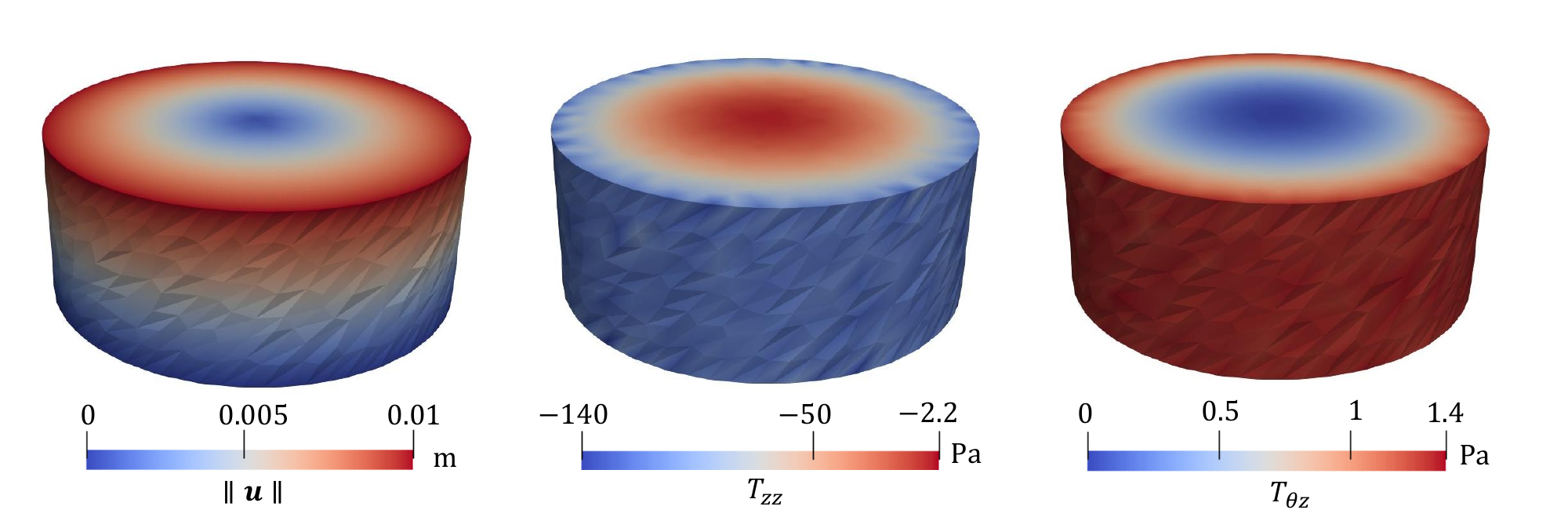}}}
    \caption{Comparison of the displacement field magnitude $\norm{\bm{u}}$ and the components of the Cauchy stress tensor $T_{zz}$ and $T_{\theta z}$ for sample $S_{24}$ at $t=5\,\,\text{s}$.}
    \label{fig: Numerical simulation}
\end{figure}

As illustrated in Fig.~\ref{fig: Comparison of results}, the torque and normal force computed from the numerical solutions closely align with both the MQLV analytical predictions and the experimental data. However, a slight discrepancy is observed between the analytical predictions and the simulations for sample $S_{24}$, likely due to the high twist rate of $400\,\,\text{rad}\,\,\text{m}^{-1}\,\,\text{s}^{-1}$. In particular, achieving convergence at high twist rates requires a very small time step, which substantially increases the complexity of the simulations and poses challenges in obtaining accurate results. Finally, in Fig.~\ref{fig: Numerical simulation}, we present an example of the full 3D simulations, showing the magnitude of the displacement field
and the components of the Cauchy stress tensor $T_{zz}$ and $T_{\theta z}$ for sample $S_{24}$ at $t=5\,\,\text{s}$.


\section{Discussion and conclusions}
\label{sec: conclusion}

\begin{figure*}[]
    {\centerline{\includegraphics[scale=1.35]{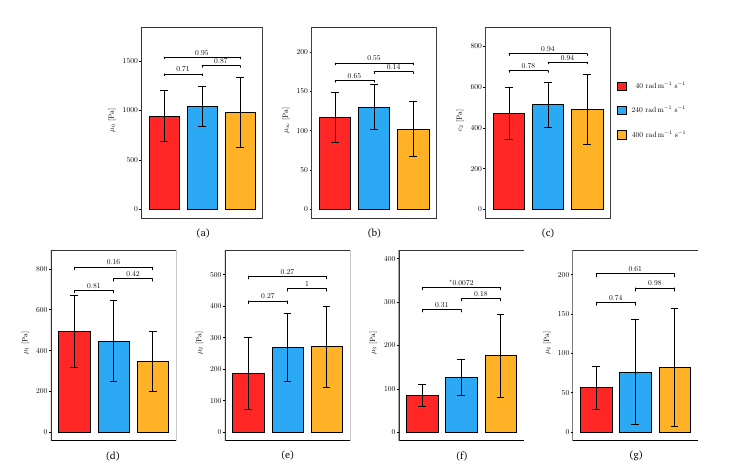}}}
    \caption{Column plots ($\text{mean} \pm \text{SD}$) of the estimated MQLV parameters (a) $\mu_0$, (b) $\mu_{\infty}$, (c) $c_2$, (d) $\mu_1$, (e) $\mu_2$, (f) $\mu_3$ and (g) $\mu_4$ for samples tested at $40\,\,\text{rad}\,\,\text{m}^{-1}\,\,\text{s}^{-1}$ (red), $240\,\,\text{rad}\,\,\text{m}^{-1}\,\,\text{s}^{-1}$ (blue) and $400\,\,\text{rad}\,\,\text{m}^{-1}\,\,\text{s}^{-1}$ (orange). Also shown are the $p$-values obtained from Tukey multiple comparisons tests, with asterisks denoting a statistically significant difference ($p<0.05$).}
    \label{fig: Statistical analysis}
\end{figure*}

\begin{figure*}[]
    {\centerline{\includegraphics[scale=1.475]{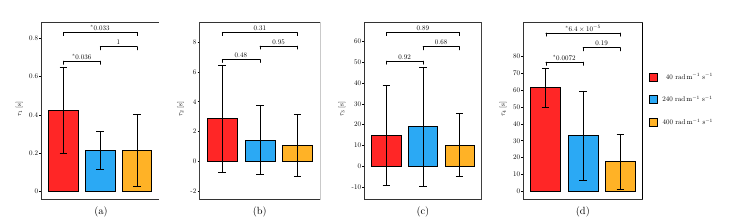}}}
    \caption{Column plots ($\text{mean} \pm \text{SD}$) of the estimated MQLV parameters (a) $\tau_1$, (b) $\tau_2$, (c) $\tau_3$ and (d) $\tau_4$ for samples tested at $40\,\,\text{rad}\,\,\text{m}^{-1}\,\,\text{s}^{-1}$ (red), $240\,\,\text{rad}\,\,\text{m}^{-1}\,\,\text{s}^{-1}$ (blue) and $400\,\,\text{rad}\,\,\text{m}^{-1}\,\,\text{s}^{-1}$ (orange). Also shown are the $p$-values obtained from Tukey multiple comparisons tests, with asterisks denoting a statistically significant difference ($p<0.05$).}
    \label{fig: Statistical analysis Tau}
\end{figure*}

To investigate the differences between the estimated MQLV parameters at the twist rates $\{40, 240, 400\}\,\,\text{rad}\,\,\text{m}^{-1}\,\,\text{s}^{-1}$, we performed Tukey multiple comparisons tests using the R (Version 4.4.2) function TukeyHSD \cite{RTukeyHSDDoc}. The column plots in Fig.~\ref{fig: Statistical analysis} show that there are no statistically significant differences between the shear moduli and second Mooney--Rivlin parameters at the three twist rates, except for the $\mu_3$ values at $40\,\,\text{rad}\,\,\text{m}^{-1}\,\,\text{s}^{-1}$ and $400\,\,\text{rad}\,\,\text{m}^{-1}\,\,\text{s}^{-1}$. By contrast, there are statistically significant differences between several of the relaxation times at the three twist rates (see Fig.~\ref{fig: Statistical analysis Tau}). 
However, the statistically significant differences for these Prony parameters may be attributed to their large standard deviations, particularly for the relaxation times. In practice, the representation of relaxation data by a Prony series is non-unique, with the Prony parameters being highly sensitive to small changes in the data \cite{fung93}. Finally, we note that the values of $\mu_0$ and $c_2$ obtained here for ovine brain tissue are in line with those reported by Balbi \textit{et al.} \cite{balbi19} for porcine brain tissue at $300\,\,\text{rad}\,\,\text{m}^{-1}\,\,\text{s}^{-1}$.

Our proposed testing protocol presents some challenges and limitations. Due to natural variations in brain size between sheep and brain tissue's highly compliant and ultra-soft nature, it was difficult to consistently and reliably prepare cylindrical samples of similar dimensions, as in Fig.~\ref{fig: Cutting protocol}. As a result, difficult-to-obtain and otherwise useful brain samples were wasted, leading to a smaller sample size. Furthermore, if the samples are not satisfactorily cylindrical (as the rotational rheometer requires), artefacts could potentially be introduced into the measured torque and normal force.

Currently, there is no consensus on the effects of temperature and post-mortem storage time on the mechanical properties of brain tissue \cite{budday20}. Therefore, it is important to note that the brains in this study were prepared and tested at room temperature ($19$--$23$\,\textdegree \text{C}) within 8 hours post-mortem.

Given the limited availability and ethical challenges associated with obtaining fresh human brains in Ireland, ovine brains were used in this work. Despite the widespread use of porcine \cite{balbi19,rashid14,sundaresh21,arbogast98} and bovine \cite{budday15,Takhounts03,bilston01} brain tissues as surrogates for characterising the mechanical properties of human brain tissue, there are relativly few studies that focus on ovine brain tissue \cite{lilley20,feng13}. 
By contrast, due to the neuroanatomical similarities between the sheep and human brains \cite{banstola22}, there is a growing body of literature that utilises the ovine model to investigate brain injuries, including strokes and epilepsy, among others \cite{lee15}. It stands to reason that ovine brain tissue, like porcine and bovine brain tissues, is an appropriate alternative to human brain tissue for our purposes. 

Despite the oulined limitations, in this paper, we devised the first experimental protocol to determine the non-linear viscoelastic properties of brain tissue in torsion. This protocol allows us to obtain two independent datasets (torque and normal force) with a single test, providing us with a much more efficient protocol compared to protocols involving multiple loading modes. The latter require a sample to be sequentially tested under different deformation modes to obtain independent datasets. Moreover, they often rely on expensive, custom-made experimental rings or multiple testing devices. Our novel protocol can be easily implemented in any commercially available rheometer and has huge potential to accurately model the non-linear viscoelastic properties of brain tissue. 

Here, we applied the protocol to study the non-linear viscoelastic behaviour of the brain in torsion at varying twist rates. This protocol has huge potential not only to study the strain-dependent relaxation of the brain but also its non-linear creep behaviour. From the theoretical viewpoint, we showed that the MQLV model provides a good fit to the experimental data and allows us to estimate the time-dependent shear modulus of an incompressible, viscoelastic, soft material such as the brain. The fitting procedure that we proposed can also be applied to compressible, viscoelastic, soft materials, whose mechanical behaviour is determined by at least two material functions.    

Finally, when coupled with bespoke finite element models such as the University College Dublin Brain Trauma Model \cite{horgan03}, the viscoelastic material parameters estimated in this study could enhance our understanding of the forces and deformations involved in traumatic brain injury and contribute to the design of improved headgear for sports such as boxing and motorsports. Our novel testing protocol also offers new insights into the mechanical behaviour of soft tissues other than the brain.


\section*{Author contributions}
\textbf{Griffen Small:} Investigation, Formal analysis, Visualization, Writing - original draft, Writing - reviewing and editing. \textbf{Francesca Ballatore:} Investigation, Software, Visualization, Writing - original draft, Writing - reviewing and editing. \textbf{Chiara Giverso:} Software, Visualization, Writing - original draft, Writing - reviewing and editing. \textbf{Valentina Balbi:} Conceptualization, Methodology, Writing - reviewing and editing.


\section*{Conflicts of interest}

There are no conflicts to declare.


\section*{Data availability}

The brain torsion data used in this article has been deposited on Mendeley Data at \href{https://doi.org/10.17632/m2jwdfgczs.1}{https://doi.org/10.17632/m2jwdfgczs.1}. The FEniCS code for the finite element simulations is available on Github at \href{https://github.com/francescaballatore/Brain_QLV_Torsion}{https://github.com/francescaballatore/Brain\_QLV\_Torsion}.


\section*{Acknowledgements}

This publication has emanated from research jointly funded by Taighde \'Eireann -- Research Ireland under grant number \\ GOIPG/2024/3552 (Griffen Small), and by the College of Science and Engineering at the University of Galway under the Millennium Fund scheme for the project “Modelling Brain Mechanics’’ (Valentina Balbi).


\section*{Appendix A}
\label{sec: appendix a}
The functions $A,B,C,D$ and $E$ appearing in the torque \eqref{Hold phase torque} and normal force \eqref{Hold phase normal force} are given by
\begin{alignat*}{3}
    &A(\lambda,\gamma)=\frac{1}{\lambda}[\lambda+1+2\gamma\left(\lambda-1\right)], \qquad &&B(\lambda,\gamma)=\frac{1}{\lambda}[\lambda+2+2\gamma\left(\lambda-2\right)], \qquad &C(\lambda)=\lambda^4+2\lambda^3+2\lambda-2, \\[5pt]
    &D(\lambda)=\lambda^4-\lambda^3-\lambda+1, \qquad &&E(\lambda)=\lambda^4-\lambda^3+2\lambda-2. \qquad &\phantom{x}  
\end{alignat*}


\bibliography{Brain_MQLV_Paper.bbl}

\begin{thebibliography}{10}
\expandafter\ifx\csname url\endcsname\relax
  \def\url#1{\texttt{#1}}\fi
\expandafter\ifx\csname urlprefix\endcsname\relax\def\urlprefix{URL }\fi
\expandafter\ifx\csname href\endcsname\relax
  \def\href#1#2{#2} \def\path#1{#1}\fi

\bibitem{budday20}
S.~Budday, T.~C. Ovaert, G.~A. Holzapfel, P.~Steinmann, E.~Kuhl, Fifty {S}hades of {B}rain: {A} {R}eview on the {M}echanical {T}esting and {M}odeling of {B}rain {T}issue, Archives of Computational Methods in Engineering 27 (2020) 1187--1230.
\newblock \href {https://doi.org/10.1007/s11831-019-09352-w} {\path{doi:10.1007/s11831-019-09352-w}}.

\bibitem{balbi19}
V.~Balbi, A.~Trotta, M.~Destrade, A.~N{\'\i}~Annaidh, Poynting effect of brain matter in torsion, Soft Matter 15~(25) (2019) 5147--5153.
\newblock \href {https://doi.org/10.1039/c9sm00131j} {\path{doi:10.1039/c9sm00131j}}.

\bibitem{destrade15}
M.~Destrade, M.~D. Gilchrist, J.~G. Murphy, B.~Rashid, G.~Saccomandi, {E}xtreme softness of brain matter in simple shear, International Journal of Non-Linear Mechanics 75 (2015) 54--58.
\newblock \href {https://doi.org/10.1016/j.ijnonlinmec.2015.02.014} {\path{doi:10.1016/j.ijnonlinmec.2015.02.014}}.

\bibitem{greiner21}
A.~Greiner, N.~Reiter, F.~Paulsen, G.~A. Holzapfel, P.~Steinmann, E.~Comellas, S.~Budday, Poro-viscoelastic effects during biomechanical testing of human brain tissue, Frontiers in Mechanical Engineering 7 (2021) 708350.
\newblock \href {https://doi.org/10.3389/fmech.2021.708350} {\path{doi:10.3389/fmech.2021.708350}}.

\bibitem{rashid14}
B.~Rashid, M.~Destrade, M.~D. Gilchrist, Mechanical characterization of brain tissue in tension at dynamic strain rates, Journal of the Mechanical Behavior of Biomedical Materials 33 (2014) 43--54.
\newblock \href {https://doi.org/10.1016/j.jmbbm.2012.07.015} {\path{doi:10.1016/j.jmbbm.2012.07.015}}.

\bibitem{rashid13}
B.~Rashid, M.~Destrade, M.~D. Gilchrist, Mechanical characterization of brain tissue in simple shear at dynamic strain rates, Journal of the Mechanical Behavior of Biomedical Materials 28 (2013) 71--85.
\newblock \href {https://doi.org/10.1016/j.jmbbm.2013.07.017} {\path{doi:10.1016/j.jmbbm.2013.07.017}}.

\bibitem{rashid12(2)}
B.~Rashid, M.~Destrade, M.~D. Gilchrist, Mechanical characterization of brain tissue in compression at dynamic strain rates, Computational Materials Science 10 (2012) 23--38.
\newblock \href {https://doi.org/10.1016/j.jmbbm.2012.01.022} {\path{doi:10.1016/j.jmbbm.2012.01.022}}.

\bibitem{budday17}
S.~Budday, G.~Sommer, J.~Haybaeck, P.~Steinmann, G.~A. Holzapfel, E.~Kuhl, Rheological characterization of human brain tissue, Acta Biomaterialia 60 (2017) 315--329.
\newblock \href {https://doi.org/10.1016/j.actbio.2017.06.024} {\path{doi:10.1016/j.actbio.2017.06.024}}.

\bibitem{kang24}
W.~Kang, L.~Wang, Y.~Fan, Viscoelastic response of gray matter and white matter brain tissues under creep and relaxation, Journal of Biomechanics 162 (2024) 111888.
\newblock \href {https://doi.org/10.1016/j.jbiomech.2023.111888} {\path{doi:10.1016/j.jbiomech.2023.111888}}.

\bibitem{GIVERSO_stress_relax}
C.~Giverso, L.~Preziosi, Modelling the compression and reorganization of cell aggregates, Mathematical Medicine and Biology: A Journal of the IMA 29~(2) (2010) 181--204.
\newblock \href {https://doi.org/10.1093/imammb/dqr008} {\path{doi:10.1093/imammb/dqr008}}.

\bibitem{Ambrosi_Preziosi_stressrelax}
D.~Ambrosi, L.~Preziosi, Cell adhesion mechanisms and stress relaxation in the mechanics of tumours, Biomechanics and Modeling in Mechanobiology 8~(5) (2009) 397--413.
\newblock \href {https://doi.org/10.1007/s10237-008-0145-y} {\path{doi:10.1007/s10237-008-0145-y}}.

\bibitem{GIVERSO_creep}
C.~Giverso, L.~Preziosi, Behavior of cell aggregates under force-controlled compression, International Journal of Non-Linear Mechanics 56 (2013) 50--55.
\newblock \href {https://doi.org/10.1016/j.ijnonlinmec.2013.05.006} {\path{doi:10.1016/j.ijnonlinmec.2013.05.006}}.

\bibitem{delingette04}
H.~Delingette, N.~Ayache, Soft tissue modeling for surgery simulation, Handbook of Numerical Analysis 12 (2004) 453--550.
\newblock \href {https://doi.org/10.1016/S1570-8659(03)12005-4} {\path{doi:10.1016/S1570-8659(03)12005-4}}.

\bibitem{ji22}
S.~Ji, M.~Ghajari, H.~Mao, R.~H. Kraft, M.~Hajiaghamemar, M.~B. Panzer, R.~Willinger, M.~D. Gilchrist, S.~Kleiven, J.~D. Stitzel, Use of brain biomechanical models for monitoring impact exposure in contact sports, Annals of Biomedical Engineering 50~(11) (2022) 1389--1408.
\newblock \href {https://doi.org/10.1007/s10439-022-02999-w} {\path{doi:10.1007/s10439-022-02999-w}}.

\bibitem{connor19}
T.~A. Connor, J.~M. Clark, J.~Jayamohan, M.~Stewart, A.~McGoldrick, C.~Williams, B.~M. Seemungal, R.~Smith, R.~Burek, M.~D. Gilchrist, Do equestrian helmets prevent concussion? {A} retrospective analysis of head injuries and helmet damage from real-world equestrian accidents, Sports Medicine 5 (2019) 1--8.
\newblock \href {https://doi.org/10.1186/s40798-019-0193-0} {\path{doi:10.1186/s40798-019-0193-0}}.

\bibitem{wineman09}
A.~Wineman, Nonlinear {V}iscoelastic {S}olids---{A} {R}eview, Mathematics and Mechanics of Solids 14~(3) (2009) 300--366.
\newblock \href {https://doi.org/10.1177/1081286509103660} {\path{doi:10.1177/1081286509103660}}.

\bibitem{drapaca07}
C.~S. Drapaca, S.~Sivaloganathan, G.~Tenti, Nonlinear constitutive laws in viscoelasticity, Mathematics and Mechanics of Solids 12~(5) (2007) 475--501.
\newblock \href {https://doi.org/10.1177/1081286506062450} {\path{doi:10.1177/1081286506062450}}.

\bibitem{anand20}
L.~Anand, S.~Govindjee, Continuum {M}echanics of {S}olids, Oxford University Press, 2020.
\newblock \href {https://doi.org/10.1093/oso/9780198864721.001.0001} {\path{doi:10.1093/oso/9780198864721.001.0001}}.

\bibitem{christensen03}
R.~M. Christensen, {T}heory of {V}iscoelasticity: {A}n {I}ntroduction, 2nd Edition, Dover Publications, 2003.

\bibitem{fung93}
Y.~C. Fung, Biomechanics: {M}echanical {P}roperties of {L}iving {T}issues, 2nd Edition, Springer Science \& Business Media, 1993.
\newblock \href {https://doi.org/10.1007/978-1-4757-2257-4} {\path{doi:10.1007/978-1-4757-2257-4}}.

\bibitem{depascalis14}
R.~De~Pascalis, D.~I. Abrahams, W.~J. Parnell, On nonlinear viscoelastic deformations: a reappraisal of {F}ung's quasi-linear viscoelastic model, Proceedings of the Royal Society A: Mathematical, Physical and Engineering Sciences 470~(2166) (2014) 20140058.
\newblock \href {https://doi.org/10.1098/rspa.2014.0058} {\path{doi:10.1098/rspa.2014.0058}}.

\bibitem{shearer20}
T.~Shearer, W.~J. Parnell, B.~Lynch, H.~R. Screen, D.~I. Abrahams, A recruitment model of tendon viscoelasticity that incorporates fibril creep and explains strain-dependent relaxation, Journal of Biomechanical Engineering 142~(7) (2020) 071003.
\newblock \href {https://doi.org/10.1115/1.4045662} {\path{doi:10.1115/1.4045662}}.

\bibitem{chatelin10}
S.~Chatelin, A.~Constantinesco, R.~Willinger, Fifty years of brain tissue mechanical testing: from \textit{in vitro} to \textit{in vivo} investigations, Biorheology 47~(5--6) (2010) 255--276.
\newblock \href {https://doi.org/10.3233/BIR-2010-0576} {\path{doi:10.3233/BIR-2010-0576}}.

\bibitem{duenwald09}
S.~E. Duenwald, R.~Vanderby, R.~S. Lakes, Viscoelastic relaxation and recovery of tendon, Annals of Biomedical Engineering 37 (2009) 1131--1140.
\newblock \href {https://doi.org/10.1007/s10439-009-9687-0} {\path{doi:10.1007/s10439-009-9687-0}}.

\bibitem{nasseri02}
S.~Nasseri, L.~E. Bilston, N.~Phan-Thien, Viscoelastic properties of pig kidney in shear, experimental results and modelling, Rheologica Acta 41~(1) (2002) 180--192.
\newblock \href {https://doi.org/10.1007/s003970200017} {\path{doi:10.1007/s003970200017}}.

\bibitem{karimi16}
A.~Karimi, M.~Haghighatnama, A.~Shojaei, M.~Navidbakhsh, A.~Motevalli~Haghi, S.~J. Adnani~Sadati, Measurement of the viscoelastic mechanical properties of the skin tissue under uniaxial loading, Proceedings of the Institution of Mechanical Engineers, Part L: Journal of Materials: Design and Applications 230~(2) (2016) 418--425.
\newblock \href {https://doi.org/10.1177/1464420715575169} {\path{doi:10.1177/1464420715575169}}.

\bibitem{flynn13}
C.~Flynn, A.~J. Taberner, P.~M.~F. Nielsen, S.~Fels, Simulating the three-dimensional deformation of \textit{in vivo} facial skin, Journal of the Mechanical Behavior of Biomedical Materials 28 (2013) 484--494.
\newblock \href {https://doi.org/10.1016/j.jmbbm.2013.03.004} {\path{doi:10.1016/j.jmbbm.2013.03.004}}.

\bibitem{macmanus19}
D.~B. MacManus, M.~Maillet, S.~O'Gorman, B.~Pierrat, J.~G. Murphy, M.~D. Gilchrist, Sex-and age-specific mechanical properties of liver tissue under dynamic loading conditions, Journal of the Mechanical Behavior of Biomedical Materials 99 (2019) 240--246.
\newblock \href {https://doi.org/10.1016/j.jmbbm.2019.07.028} {\path{doi:10.1016/j.jmbbm.2019.07.028}}.

\bibitem{sundaresh22}
S.~N. Sundaresh, J.~D. Finan, B.~S. Elkin, A.~V. Basilio, G.~M. McKhann, B.~Morrison~III, {R}egion-{D}ependent {V}iscoelastic {P}roperties of {H}uman {B}rain {T}issue {U}nder {L}arge {D}eformations, Annals of Biomedical Engineering 50~(11) (2022) 1452--1460.
\newblock \href {https://doi.org/10.1007/s10439-022-02910-7} {\path{doi:10.1007/s10439-022-02910-7}}.

\bibitem{sundaresh21}
S.~N. Sundaresh, J.~D. Finan, B.~S. Elkin, C.~Lee, J.~Xiao, B.~Morrison~III, Viscoelastic characterization of porcine brain tissue mechanical properties under indentation loading, Brain Multiphysics 2 (2021) 100041.
\newblock \href {https://doi.org/10.1016/j.brain.2021.100041} {\path{doi:10.1016/j.brain.2021.100041}}.

\bibitem{macmanus20}
D.~B. MacManus, A.~Menichetti, B.~Depreitere, N.~Famaey, J.~Vander~Sloten, M.~Gilchrist, Towards animal surrogates for characterising large strain dynamic mechanical properties of human brain tissue, Brain Multiphysics 1 (2020) 100018.
\newblock \href {https://doi.org/10.1016/j.brain.2020.100018} {\path{doi:10.1016/j.brain.2020.100018}}.

\bibitem{hosseini19}
M.~Hosseini-Farid, A.~Rezaei, A.~Eslaminejad, M.~Ramzanpour, M.~Ziejewski, G.~Karami, Instantaneous and equilibrium responses of the brain tissue by stress relaxation and quasi-linear viscoelasticity theory, Scientia Iranica 26~(4) (2019) 2047--2056.
\newblock \href {https://doi.org/10.24200/sci.2019.21314} {\path{doi:10.24200/sci.2019.21314}}.

\bibitem{sahoo14}
D.~Sahoo, C.~Deck, R.~Willinger, Development and validation of an advanced anisotropic visco-hyperelastic human brain {FE} model, Journal of the Mechanical Behavior of Biomedical Materials 33 (2014) 24--42.
\newblock \href {https://doi.org/10.1016/j.jmbbm.2013.08.022} {\path{doi:10.1016/j.jmbbm.2013.08.022}}.

\bibitem{chatelin13}
S.~Chatelin, C.~Deck, R.~Willinger, An anisotropic viscous hyperelastic constitutive law for brain material finite-element modeling, Journal of Biorheology 27 (2013) 26--37.
\newblock \href {https://doi.org/10.1007/s12573-012-0055-6} {\path{doi:10.1007/s12573-012-0055-6}}.

\bibitem{daphalapurkar23}
N.~Daphalapurkar, J.~Riglin, A.~Mohan, J.~Harris, J.~Bernardin, Quasi-dynamic breathing model of the lung incorporating viscoelasticity of the lung tissue, International Journal for Numerical Methods in Biomedical Engineering 39~(8) (2023) e3744.
\newblock \href {https://doi.org/10.1002/cnm.3744} {\path{doi:10.1002/cnm.3744}}.

\bibitem{zhang18}
Z.~H. Zhang, M.~X. Pan, J.~T. Cai, J.~D. Weiland, K.~Chen, Viscoelastic properties of the posterior eye of normal subjects, patients with age-related macular degeneration, and pigs, Journal of Biomedical Materials Research Part A 106~(8) (2018) 2151--2157.
\newblock \href {https://doi.org/10.1002/jbm.a.36417} {\path{doi:10.1002/jbm.a.36417}}.

\bibitem{rycman21}
A.~Rycman, S.~McLachlin, D.~S. Cronin, A hyper-viscoelastic continuum-level finite element model of the spinal cord assessed for transverse indentation and impact loading, Frontiers in Bioengineering and Biotechnology 9 (2021) 693120.
\newblock \href {https://doi.org/10.3389/fbioe.2021.693120} {\path{doi:10.3389/fbioe.2021.693120}}.

\bibitem{yu20}
J.~Yu, N.~Manouchehri, S.~Yamamoto, B.~K. Kwon, T.~R. Oxland, Mechanical properties of spinal cord grey matter and white matter in confined compression, Journal of the Mechanical Behavior of Biomedical Materials 112 (2020) 104044.
\newblock \href {https://doi.org/10.1016/j.jmbbm.2020.104044} {\path{doi:10.1016/j.jmbbm.2020.104044}}.

\bibitem{jannesar16}
S.~Jannesar, B.~Nadler, C.~J. Sparrey, The transverse isotropy of spinal cord white matter under dynamic load, Journal of Biomechanical Engineering 138~(9) (2016) 091004.
\newblock \href {https://doi.org/10.1115/1.4034171} {\path{doi:10.1115/1.4034171}}.

\bibitem{helisaz24}
H.~Helisaz, E.~Belanger, P.~Black, M.~Bacca, M.~Chiao, Quantifying the impact of cancer on the viscoelastic properties of the prostate gland using a quasi-linear viscoelastic model, Acta Biomaterialia 173 (2024) 184--198.
\newblock \href {https://doi.org/10.1016/j.actbio.2023.11.002} {\path{doi:10.1016/j.actbio.2023.11.002}}.

\bibitem{motallebzadeh13}
H.~Motallebzadeh, M.~Charlebois, W.~R.~J. Funnell, A non-linear viscoelastic model for the tympanic membrane, The Journal of the Acoustical Society of America 134~(6) (2013) 4427--4434.
\newblock \href {https://doi.org/10.1121/1.4828831} {\path{doi:10.1121/1.4828831}}.

\bibitem{yang06}
W.~Yang, T.~C. Fung, K.~S. Chian, C.~K. Chong, Investigations of the viscoelasticity of esophageal tissue using incremental stress-relaxation test and cyclic extension test, Journal of Mechanics in Medicine and Biology 6~(03) (2006) 261--272.
\newblock \href {https://doi.org/10.1142/S0219519406001984} {\path{doi:10.1142/S0219519406001984}}.

\bibitem{huyghe91}
J.~M. Huyghe, D.~H. Van~Campen, T.~Arts, R.~M. Heethaar, The constitutive behaviour of passive heart muscle tissue: {A} quasi-linear viscoelastic formulation, Journal of Biomechanics 24~(9) (1991) 841--849.
\newblock \href {https://doi.org/10.1016/0021-9290(91)90309-b} {\path{doi:10.1016/0021-9290(91)90309-b}}.

\bibitem{criscenti15}
G.~Criscenti, C.~De~Maria, E.~Sebastiani, M.~Tei, G.~Placella, A.~Speziali, G.~Vozzi, G.~Cerulli, Quasi-linear viscoelastic properties of the human medial patello-femoral ligament, Journal of Biomechanics 48~(16) (2015) 4297--4302.
\newblock \href {https://doi.org/10.1016/j.jbiomech.2015.10.042} {\path{doi:10.1016/j.jbiomech.2015.10.042}}.

\bibitem{abramowitch04}
S.~D. Abramowitch, S.~L. Woo, An improved method to analyze the stress relaxation of ligaments following a finite ramp time based on the quasi-linear viscoelastic theory, Journal of Biomedical Engineering 126~(1) (2004) 92--97.
\newblock \href {https://doi.org/10.1115/1.1645528} {\path{doi:10.1115/1.1645528}}.

\bibitem{funk00}
J.~R. Funk, G.~W. Hall, J.~R. Crandall, W.~D. Pilkey, Linear and quasi-linear viscoelastic characterization of ankle ligaments, Journal of Biomechanical Engineering 122~(1) (2000) 15--22.
\newblock \href {https://doi.org/10.1115/1.429623} {\path{doi:10.1115/1.429623}}.

\bibitem{bah20}
I.~Bah, N.~R.~J. Fernandes, R.~L. Chimenti, J.~Ketz, A.~S. Flemister, M.~R. Buckley, Tensile mechanical changes in the {A}chilles tendon due to insertional {A}chilles tendinopathy, Journal of the Mechanical Behavior of Biomedical Materials 112 (2020) 104031.
\newblock \href {https://doi.org/10.1016/j.jmbbm.2020.104031} {\path{doi:10.1016/j.jmbbm.2020.104031}}.

\bibitem{springhetti18}
R.~Springhetti, N.~S. Selyutina, Viscoelastic modeling of articular cartilage under impact loading, Meccanica 53~(3) (2018) 519--530.
\newblock \href {https://doi.org/10.1007/s11012-017-0717-y} {\path{doi:10.1007/s11012-017-0717-y}}.

\bibitem{selyutina15}
N.~S. Selyutina, I.~I. Argatov, G.~S. Mishuris, On application of {F}ung's quasi-linear viscoelastic model to modeling of impact experiment for articular cartilage, Mechanics Research Communications 67 (2015) 24--30.
\newblock \href {https://doi.org/10.1016/j.mechrescom.2015.04.003} {\path{doi:10.1016/j.mechrescom.2015.04.003}}.

\bibitem{giudici23}
A.~Giudici, K.~W.~F. van~der Laan, M.~M. van~der Bruggen, S.~Parikh, E.~Berends, S.~Foulquier, T.~Delhaas, K.~D. Reesink, B.~Spronck, Constituent-based quasi-linear viscoelasticity: a revised quasi-linear modelling framework to capture nonlinear viscoelasticity in arteries, Biomechanics and Modeling in Mechanobiology 22~(5) (2023) 1607--1623.
\newblock \href {https://doi.org/10.1007/s10237-023-01711-8} {\path{doi:10.1007/s10237-023-01711-8}}.

\bibitem{dadgar21}
F.~Dadgar-Rad, N.~Firouzi, Time-dependent response of incompressible membranes based on quasi-linear viscoelasticity theory, International Journal of Applied Mechanics 13~(03) (2021) 2150036.
\newblock \href {https://doi.org/10.1142/S1758825121500368} {\path{doi:10.1142/S1758825121500368}}.

\bibitem{depascalis15}
R.~De~Pascalis, D.~I. Abrahams, W.~J. Parnell, Simple shear of a compressible quasilinear viscoelastic material, International Journal of Engineering Science 88 (2015) 64--72.
\newblock \href {https://doi.org/10.1016/j.ijengsci.2014.11.011} {\path{doi:10.1016/j.ijengsci.2014.11.011}}.

\bibitem{righi21}
M.~Righi, V.~Balbi, Foundations of viscoelasticity and application to soft tissue mechanics, in: J.~M{\'a}lek, E.~S{\"u}li (Eds.), Modeling Biomaterials, Springer, 2021, Ch.~3, pp. 71--103.
\newblock \href {https://doi.org/10.1007/978-3-030-88084-2_3} {\path{doi:10.1007/978-3-030-88084-2_3}}.

\bibitem{balbi23}
V.~Balbi, T.~Shearer, W.~J. Parnell, Tensor decomposition for modified quasi-linear viscoelastic models: {T}owards a fully non-linear theory, Mathematics and Mechanics of Solid 0~(0) (2023) 1--25.
\newblock \href {https://doi.org/10.1177/10812865231165232} {\path{doi:10.1177/10812865231165232}}.

\bibitem{balbi18}
V.~Balbi, T.~Shearer, W.~J. Parnell, A modified formulation of quasi-linear viscoelasticity for transversely isotropic materials under finite deformation, Proceedings of the Royal Society A: Mathematical, Physical and Engineering Sciences 474~(2217) (2018) 20180231.
\newblock \href {https://doi.org/10.1098/rspa.2018.0231} {\path{doi:10.1098/rspa.2018.0231}}.

\bibitem{depascalis18}
R.~De~Pascalis, W.~J. Parnell, D.~I. Abrahams, T.~Shearer, D.~M. Daly, D.~Grundy, The inflation of viscoelastic balloons and hollow viscera, Proceedings of the Royal Society A: Mathematical, Physical and Engineering Sciences 474~(2218) (2018) 20180102.
\newblock \href {https://doi.org/10.1098/rspa.2018.0102} {\path{doi:10.1098/rspa.2018.0102}}.

\bibitem{macmanus17}
D.~B. MacManus, B.~Pierrat, J.~G. Murphy, M.~D. Gilchrist, Region and species dependent mechanical properties of adolescent and young adult brain tissue, Scientific Reports 7~(1) (2017) 13729.
\newblock \href {https://doi.org/10.1038/s41598-017-13727-z} {\path{doi:10.1038/s41598-017-13727-z}}.

\bibitem{menichetti20}
A.~Menichetti, D.~B. MacManus, M.~D. Gilchrist, B.~Depreitere, J.~Vander~Sloten, N.~Famaey, Regional characterization of the dynamic mechanical properties of human brain tissue by microindentation, International Journal of Engineering Science 155 (2020) 103355.
\newblock \href {https://doi.org/10.1016/j.ijengsci.2020.103355} {\path{doi:10.1016/j.ijengsci.2020.103355}}.

\bibitem{garo07}
A.~Garo, M.~Hrapko, J.~A.~W. Van~Dommelen, G.~W.~M. Peters, Towards a reliable characterisation of the mechanical behaviour of brain tissue: the effects of post-mortem time and sample preparation, Biorheology 44~(1) (2007) 51--58.
\newblock \href {https://doi.org/10.1177/0006355X2007044001003} {\path{doi:10.1177/0006355X2007044001003}}.

\bibitem{rashid12(3)}
B.~Rashid, M.~Destrade, M.~D. Gilchrist, Temperature effects on brain tissue in compression, Journal of the Mechanical Behavior of Biomedical Materials 14 (2012) 113--118.
\newblock \href {https://doi.org/10.1016/j.jmbbm.2012.04.005} {\path{doi:10.1016/j.jmbbm.2012.04.005}}.

\bibitem{rashid13(2)}
B.~Rashid, M.~Destrade, M.~D. Gilchrist, Influence of preservation temperature on the measured mechanical properties of brain tissue, Journal of Biomechanics 46~(7) (2013) 1276--1281.
\newblock \href {https://doi.org/10.1016/j.jbiomech.2013.02.014} {\path{doi:10.1016/j.jbiomech.2013.02.014}}.

\bibitem{arbogast98}
K.~B. Arbogast, S.~S. Margulies, Material characterization of the brainstem from oscillatory shear tests, Journal of Biomechanics 31~(9) (1998) 801--807.
\newblock \href {https://doi.org/10.1016/S0021-9290(98)00068-2} {\path{doi:10.1016/S0021-9290(98)00068-2}}.

\bibitem{rashid12}
B.~Rashid, M.~Destrade, M.~D. Gilchrist, Inhomogeneous deformation of brain tissue during tension tests, Computational Materials Science 64 (2012) 295--300.
\newblock \href {https://doi.org/10.1016/j.commatsci.2012.05.030} {\path{doi:10.1016/j.commatsci.2012.05.030}}.

\bibitem{destrade23}
M.~Destrade, Y.~Du, J.~Blackwell, N.~Colgan, V.~Balbi, {C}anceling the elastic {P}oynting effect with geometry, Physical Review E 107~(5) (2023) L053001.
\newblock \href {https://doi.org/10.1103/PhysRevE.107.L053001} {\path{doi:10.1103/PhysRevE.107.L053001}}.

\bibitem{Britishstandards12}
{British Standards}, {P}hysical {T}esting of {R}ubber, Tech. Rep. BS 903-0:2012, British Standards Institution (2012).

\bibitem{yan21}
S.~Yan, D.~Jia, Y.~Yu, L.~Wang, Y.~Qiu, Q.~Wan, Novel strategies for parameter fitting procedure of the {O}gden hyperfoam model under shear condition, European Journal of Mechanics A / Solids 86 (2021) 104154.
\newblock \href {https://doi.org/10.1016/j.euromechsol.2020.104154} {\path{doi:10.1016/j.euromechsol.2020.104154}}.

\bibitem{rivlin49}
R.~S. Rivlin, Large elastic deformations of isotropic materials {VI}. {F}urther results in the theory of torsion, shear and flexure, Philosophical Transactions of the Royal Society of London. Series A, Mathematical and Physical Sciences 242~(845) (1949) 173--195.
\newblock \href {https://doi.org/10.1098/rsta.1949.0009} {\path{doi:10.1098/rsta.1949.0009}}.

\bibitem{poynting1909}
J.~H. Poynting, On pressure perpendicular to the shear planes in finite pure shears, and on the lengthening of loaded wires when twisted, Proceedings of the Royal Society of London. Series A, Containing Papers of a Mathematical and Physical Character 82~(557) (1909) 546--559.
\newblock \href {https://doi.org/10.1098/rspa.1909.0059} {\path{doi:10.1098/rspa.1909.0059}}.

\bibitem{narayan12}
S.~P.~A. Narayan, J.~M. Krishnan, A.~P. Deshpande, K.~R. Rajagopal, Nonlinear viscoelastic response of asphalt binders: {A}n experimental study of the relaxation of torque and normal force in torsion, Mechanics Research Communications 43 (2012) 66--74.
\newblock \href {https://doi.org/10.1016/j.mechrescom.2012.02.012} {\path{doi:10.1016/j.mechrescom.2012.02.012}}.

\bibitem{MATLABfilterDoc}
{The MathWorks Inc.}, \href{https://mathworks.com/help/signal/ref/sgolayfilt.html}{Documentation for sgolayfilt}.
\newline\urlprefix\url{https://mathworks.com/help/signal/ref/sgolayfilt.html}

\bibitem{ciarletta14}
P.~Ciarletta, M.~Destrade, Torsion instability of soft solid cylinders, The IMA Journal of Applied Mathematics 79~(5) (2014) 804--819.
\newblock \href {https://doi.org/10.1093/imamat/hxt052} {\path{doi:10.1093/imamat/hxt052}}.

\bibitem{mooney40}
M.~Mooney, A theory of large elastic deformation, Journal of Applied Physics 11~(9) (1940) 582--592.
\newblock \href {https://doi.org/10.1063/1.1712836} {\path{doi:10.1063/1.1712836}}.

\bibitem{rivlin48}
R.~S. Rivlin, Large elastic deformations of isotropic materials {IV}. {F}urther developments of the general theory, Philosophical Transactions of the Royal Society of London. Series A, Mathematical and Physical Sciences 241~(835) (1948) 379--397.
\newblock \href {https://doi.org/10.1098/rsta.1948.0024} {\path{doi:10.1098/rsta.1948.0024}}.

\bibitem{holzapfel00}
G.~A. Holzapfel, {N}onlinear {S}olid {M}echanics: {A} {C}ontinuum {A}pproach for {E}ngineering, John Wiley {\&} Sons Ltd., 2000.

\bibitem{ogden97}
R.~W. Ogden, {N}on-{L}inear {E}lastic {D}eformations, Dover Publications, 1997.

\bibitem{morrison23}
O.~Morrison, M.~Destrade, B.~B. Tripathi, An atlas of the heterogeneous viscoelastic brain with local power-law attenuation synthesised using {P}rony-series, Acta Biomaterialia 169 (2023) 66--87.
\newblock \href {https://doi.org/10.1016/j.actbio.2023.07.040} {\path{doi:10.1016/j.actbio.2023.07.040}}.

\bibitem{gilchrist13}
M.~D. Gilchrist, B.~Rashid, J.~G. Murphy, G.~Saccomandi, Quasi-static deformations of biological soft tissue, Mathematics and Mechanics of Solids 18~(6) (2013) 622--633.
\newblock \href {https://doi.org/10.1177/1081286513485770} {\path{doi:10.1177/1081286513485770}}.

\bibitem{MATLABfminconDoc}
{The MathWorks Inc.}, \href{https://mathworks.com/help/optim/ug/fmincon.html}{Documentation for fmincon}.
\newline\urlprefix\url{https://mathworks.com/help/optim/ug/fmincon.html}

\bibitem{anssari22}
A.~Anssari-Benam, M.~Destrade, G.~Saccomandi, Modelling brain tissue elasticity with the {O}gden model and an alternative family of constitutive models, Philosophical Transactions of the Royal Society A: Mathematical, Physical and Engineering Sciences 380~(2234) (2022) 20210325.
\newblock \href {https://doi.org/10.1098/rsta.2021.0325} {\path{doi:10.1098/rsta.2021.0325}}.

\bibitem{destrade17}
M.~Destrade, G.~Saccomandi, I.~Sgura, Methodical fitting for mathematical models of rubber-like materials, Proceedings of the Royal Society A: Mathematical, Physical and Engineering Sciences 473~(2198) (2017) 20160811.
\newblock \href {https://doi.org/10.1098/rspa.2016.0811} {\path{doi:10.1098/rspa.2016.0811}}.

\bibitem{alnaes15}
M.~Aln{\ae}s, J.~Blechta, J.~Hake, A.~Johansson, B.~Kehlet, A.~Logg, C.~Richardson, J.~Ring, M.~E. Rognes, G.~N. Wells, The {FE}ni{CS} project version 1.5, Archive of Numerical Software 3~(100) (2015) 9--23.
\newblock \href {https://doi.org/10.11588/ans.2015.100.20553} {\path{doi:10.11588/ans.2015.100.20553}}.

\bibitem{logg12}
A.~Logg, K.~A. Mardal, G.~Wells, {A}utomated {S}olution of {D}ifferential {E}quations by the {F}inite {E}lement {M}ethod, Springer, 2012.
\newblock \href {https://doi.org/10.1007/978-3-642-23099-8} {\path{doi:10.1007/978-3-642-23099-8}}.

\bibitem{Paraview}
U.~Ayachit, {T}he {P}ara{V}iew {G}uide: {A} {P}arallel {V}isualization {A}pplication, Kitware Inc., 2015.

\bibitem{RTukeyHSDDoc}
{R Core Team}, \href{https://www.rdocumentation.org/packages/stats/versions/3.6.2/topics/TukeyHSD}{Documentation for tukeyhsd}.
\newline\urlprefix\url{https://www.rdocumentation.org/packages/stats/versions/3.6.2/topics/TukeyHSD}

\bibitem{budday15}
S.~Budday, R.~Nay, R.~De~Rooij, P.~Steinmann, T.~Wyrobek, T.~C. Ovaert, E.~Kuhl, Mechanical properties of gray and white matter brain tissue by indentation, Journal of the Mechanical Behavior of Biomedical Materials 46 (2015) 318--330.
\newblock \href {https://doi.org/10.1016/j.jmbbm.2015.02.024} {\path{doi:10.1016/j.jmbbm.2015.02.024}}.

\bibitem{Takhounts03}
E.~G. Takhounts, J.~R. Crandall, K.~Darvish, On the importance of nonlinearity of brain tissue under large deformations, Stapp Car Crash Journal 47 (2003) 107--134.
\newblock \href {https://doi.org/10.4271/2003-22-0005} {\path{doi:10.4271/2003-22-0005}}.

\bibitem{bilston01}
L.~E. Bilston, Z.~Liu, N.~Phan-Thien, Large strain behaviour of brain tissue in shear: some experimental data and differential constitutive model, Biorheology 38~(4) (2001) 335--345.
\newblock \href {https://doi.org/10.1177/0006355X2001038004004} {\path{doi:10.1177/0006355X2001038004004}}.

\bibitem{lilley20}
R.~Lilley, A.~Reynaud, P.~D. Docherty, N.~Smith, N.~Kabaliuk, Rheological experimentation to investigate history dependent viscoelastic properties of \textit{ex-vivo} ovine brain tissue, IFAC-PapersOnLine 53~(2) (2020) 16275--16280.
\newblock \href {https://doi.org/10.1016/j.ifacol.2020.12.623} {\path{doi:10.1016/j.ifacol.2020.12.623}}.

\bibitem{feng13}
Y.~Feng, R.~J. Okamoto, R.~Namani, G.~M. Genin, P.~V. Bayly, Measurements of mechanical anisotropy in brain tissue and implications for transversely isotropic material models of white matter, Journal of the Mechanical Behavior of Biomedical Materials 23 (2013) 117--132.
\newblock \href {https://doi.org/10.1016/j.jmbbm.2013.04.007} {\path{doi:10.1016/j.jmbbm.2013.04.007}}.

\bibitem{banstola22}
A.~Banstola, J.~N.~J. Reynolds, Mapping sheep to human brain: {T}he need for a sheep brain atlas, Frontiers in Veterinary Science 9 (2022) 961413.
\newblock \href {https://doi.org/10.3389/fvets.2022.961413} {\path{doi:10.3389/fvets.2022.961413}}.

\bibitem{lee15}
W.~Lee, S.~D. Lee, M.~Y. Park, L.~Foley, E.~Purcell-Estabrook, H.~Kim, S.~S. Yoo, Functional and diffusion tensor magnetic resonance imaging of the sheep brain, BMC Veterinary Research 11 (2015) 1--8.
\newblock \href {https://doi.org/10.1186/s12917-015-0581-8} {\path{doi:10.1186/s12917-015-0581-8}}.

\bibitem{horgan03}
T.~J. Horgan, M.~D. Gilchrist, The creation of three-dimensional finite element models for simulating head impact biomechanics, International Journal of Crashworthiness 8~(4) (2003) 353--366.
\newblock \href {https://doi.org/10.1533/ijcr.2003.0243} {\path{doi:10.1533/ijcr.2003.0243}}.

\end{thebibliography}
\bibliographystyle{elsarticle-num}


\end{document}